\documentclass[11pt,leqno]{article}
\usepackage{amsmath,amssymb,amsfonts,amsthm,mathrsfs,verbatim} 
\usepackage{array}
\usepackage[noend]{algpseudocode}
\usepackage{setspace}
\usepackage{authblk}
\usepackage{caption}
\usepackage{geometry}
\usepackage{graphics} 
\usepackage{graphicx}  
\usepackage{lscape}
\usepackage{lscape}
\usepackage{natbib}
\usepackage{hyperref}
\usepackage{makeidx}
\usepackage{color}
\usepackage{bm}
\usepackage{soul}
\usepackage{tikz}
\usepackage{cancel}
\usepackage{subcaption}

\newtheorem{theorem}{Theorem}[section]

\newtheorem{algorithm}[theorem]{Algorithm}

\theoremstyle{definition}

\title{Probabilistic Scenario-Based Assessment of National Food Security Risks with Application to Egypt and Ethiopia}

\date{ }
\author[a]{P. Koundouri}
\author[b]{G. I. Papayiannis\footnote{Corresponding author: \url{gpapagiannis@aueb.gr}; \,\,\, \url{g.papagiannis@hna.gr}}}
\author[c]{A. Vassilopoulos}
\author[d]{A. N. Yannacopoulos}

\affil[a]{\footnotesize School of Economics, Athens University of Economics and Business, Athens, GR 10434; 

Department of Technology, Management and Economics, Technical University of Denmark, DK 2800; 

ReSEES Laboratory, Athens University of Economics and Business, Athens, GR 10434; 

Sustainable Development Unit, ATHENA RC, Marousi, GR 15125; 

Sustainable Development Solutions Network-Europe, Paris, FR 75009; 

Academia Europaea, London, UK WC1E7HU

\mbox{  }}

\affil[b]{\footnotesize Section of Mathematics, Department of Naval Sciences, Hellenic Naval Academy, Piraeus, GR 18539; 

Stochastic Modelling and Applications Laboratory, Athens University of Economics \& Business, Athens, GR 10434

\mbox{  }}

\affil[c]{\footnotesize Department of Agricultural Economics \& Development, School of Applied Economics and Social Sciences, Agricultural University of Athens, Athens, GR 11855

\mbox{  }}

\affil[d]{\footnotesize Department of Statistics, Athens University of Economics \& Business, Athens, GR 10434;

Stochastic Modelling and Applications Laboratory, Athens University of Economics \& Business, Athens, GR 10434}

\graphicspath{ {Figures/} } 

\begin{document}

\maketitle

\begin{abstract}
This study presents a novel approach to assessing food security risks at the national level, employing a probabilistic scenario-based framework that integrates both Shared Socioeconomic Pathways (SSP) and Representative Concentration Pathways (RCP). This innovative method allows each scenario, encompassing socio-economic and climate factors, to be treated as a model capable of generating diverse trajectories.  This approach offers a more dynamic understanding of food security risks under varying future conditions. The paper details the methodologies employed, showcasing their applicability through a focused analysis of food security challenges in Egypt and Ethiopia, and underscores the importance of considering a spectrum of socio-economic and climatic factors in national food security assessments.
\end{abstract}


\section{Introduction}\label{sec-1}

In the era of rapid global change, understanding the intricate relationship between population dynamics, climate change, and food security has become more critical than ever. Evolving demographic trends, coupled with climatic shifts, are shaping the future landscape of food availability and access, calling for a comprehensive analysis that delves into the multifaceted nature of food security, acknowledging that it is not solely an issue of production but also one deeply intertwined with socioeconomic and environmental factors. The global population is projected to reach nearly 10 billion by 2050, introducing profound challenges and demands on the food supply systems. Population growth, urbanization, and demographic transitions significantly influence food demand patterns, dietary preferences, and the overall caloric requirements of nations. This demographic shift, while a central element, does not exist in isolation. It interacts with and is influenced by broader socioeconomic changes, including economic development, technological advancements, and policy frameworks. These factors collectively dictate the availability, accessibility, and utilization of food resources, thereby shaping national food security profiles. In addition, climate change emerges as a formidable force, altering the fundamentals of food production. Shifts in temperature, precipitation patterns, and increased frequency of extreme weather events have profound implications for agricultural productivity, food prices, and ultimately, the stability of food supply chains. The growing body of literature underscores the profound impact of climatic shifts on food systems worldwide. Discussions focus on the need for effective measures to tackle food security challenges posed by climate change and extreme events. However, there is a view that stringent climate mitigation policies might have more adverse effects than climate change itself on global hunger and food consumption. It is crucial to understand how climatic factors interact with population dynamics to anticipate future food security challenges. Methodological frameworks, scenario-based approaches, and global model applications are recognized for their effectiveness in providing accurate food security risk assessments, considering uncertainties in factors like demographics and economic growth. While much research assesses global food security risks, national food security and the role of population dynamics are also key concerns, with some studies focusing on regional or national levels.

 In this study, our focus is on modeling, quantifying, and predicting food security risks at the national level. We adopt a bottom-up probabilistic modeling approach, emphasizing a critical determinant of food security: population and its structure. We generate probabilistic population projections within the framework of plausible socioeconomic scenarios, particularly under the Shared Socioeconomic Pathways (SSP) framework \cite{lutz2014, lutz2018}. These projections are then integrated into a global macroeconomic model to produce SSP-compatible forecasts for essential drivers of national economies, such as GDP and labor. Utilizing these probabilistic models, we aim to contribute towards a more holistic assessment of national food security risks.  Our approach offers a multi-dimensional view, considering how socio-economic development paths might mitigate or exacerbate the impacts of climate change on food security. Given the complexities outlined, we recognize the  need to integrate both socioeconomic and climatic considerations adopting a comprehensive framework that blends the socioeconomic scenarios with climate scenarios under the Representative Concentration Pathways (RCP) framework \cite{meins2020}. This integration is crucial as it allows us to capture the interplay between human socio-economic development and climatic changes, offering a more holistic understanding of food security risks.

Accurate long-term forecasts of caloric requirements and  total caloric content of food available to the population are crucial in assessing future food balance. By quantifying the disparities between these two metrics over an extended time horizon, policymakers are afforded the opportunity to implement long-term strategies to avoid food risks. These strategies can encompass a diverse portfolio, including restructuring agricultural production, negotiating international trade agreements, and adopting scientific advances or modern technologies. Such measures can effectively mitigate potential future food security risks. For these predictions to be reliable, they must be grounded in probabilistic modeling, which provides the full distribution of relevant variables rather than mere point estimates. This approach offers a more complete picture of future trends and their probabilities, adding a depth of understanding that point estimates lack. Additionally, it is vital to incorporate model uncertainty into these predictions. This aspect is particularly critical in long-term forecasting, where the impact of stochastic factors can only be partially anticipated. Acknowledging and accounting for this uncertainty ensures a more robust and realistic assessment of potential future scenarios in food security.
 
To this end, we introduce a food security risk index designed to effectively quantify disparities between caloric requirements and  total available caloric content of food within and across the devised socio-economic and climate scenarios. Our study offers a specialized framework for assessing food security risks at the national level, employing a bottom-up modeling approach. This builds upon the probabilistic extension of the population model as outlined in \cite{raftery2012}. The model, based on a Bayesian hierarchical modeling procedure (BHM) as seen in \cite{gelman1995}, provides detailed projections of the world's future population and its age structure by country, under various probabilistic scenarios. Unlike traditional models that offer a single point prediction, our approach yields a distribution of possible population counts for a specific time instant. Our first major contribution is the modification of this population model structure, allowing us to generate probabilistic population scenarios within the SSP framework. This enables the creation of population trajectories with predefined SSP statuses. The second key contribution is integrating these population projections with a global macroeconomic model, specifically MaGE \cite{foure2013,foure2021}, to produce SSP-compatible projections for vital economic drivers such as GDP and labor. Furthermore, we blend this probabilistic scenario generation scheme with climate scenarios under the Representative Concentration Pathways (RCP) framework \cite{meins2020}. This comprehensive approach provides a robust framework for policy assessment, considering uncertainties in future predictions. Crucially, our probabilistic modeling framework transforms the qualitative narratives of SSP/RCP scenarios into actionable quantitative scenarios, making them valuable tools for quantitative policy and decision-making (refer to Section \ref{POP-SCENARIOS} for details).

Our modeling process encompasses several steps. Initially, using the projected population structures (by age and gender), and minimum caloric intake requirements per age group and gender (as determined by nutrition experts), we estimate the minimum caloric intake necessary for subsistence at the national level. Subsequently, we assess the capacities of national food systems under their existing patterns, forming models based on fundamental socioeconomic drivers (population, income, labor) and climate factors (temperature, precipitation). These models, calibrated with historical data, are then combined with projected population and other socioeconomic drivers to provide probabilistic forecasts of food system capacities up to 2050. Building on these quantities, we introduce a novel food security risk indicator. This index offers several advantages: (a) it requires less detailed data for computation compared to established food security risk indicators (discussed in Section \ref{model}), (b) it is computationally efficient, (c) it projects future food security scenarios, and (d) it incorporates the impact on natural resources, such as national water sources. The foundation of this index is the concept of convex risk measures (referenced in \cite{detlefsen2005, follmer2002convex, frittelli2002}), which effectively mitigate the effects of model uncertainty, providing robust risk assessments within and across various probabilistic scenarios. These assessments are based on the concept of Fréchet utilities or risk measures (\cite{papayiannis2018}, \cite{petrakou2022}), which are particularly valuable for long-term policy planning and implementation, crucial in scenarios where the exact future circumstances have yet to be fully determined.

The paper is structured as follows: Section \ref{POP-MODELS} details the probabilistic extension of the population model and its adaptation to align with the Shared Socioeconomic Pathways (SSP) framework. This section also introduces the set of socio-economic and climate scenarios that are central to our analysis. Subsequently, Section \ref{model} delves into the modeling approaches and considerations for determining the minimum caloric intake and evaluating the capacity of food systems at the national level. In addition, this section presents our newly developed food security risk index, along with a comprehensive discussion and comparative analysis with established indicators in the field. Finally, Section \ref{sec-4} applies our proposed approach to specific case studies: Egypt and Ethiopia. Utilizing data from 1990 to 2018, we offer projections up to the year 2050, examining both within and across various SSP-RCP scenario frameworks. This practical implementation demonstrates the applicability and effectiveness of our methodology in real-world settings.

\section{Socio-Economic and Climatic Scenario Modeling}\label{POP-MODELS}

From a modeling perspective, the literature explores various methodological approaches to food security, as outlined in \cite{krishnamurthy2014methodological}, and emphasizes the importance of accurately modeling crucial food demand factors such as demographics and economic growth, given their inherent uncertainties \cite{valin2014future}. Scenario-based approaches \cite{hasegawa2015scenarios, molotoks2021impacts, van2021meta} and global model applications \cite{van2020modelling}, as well as integrative modeling efforts \cite{muller2020modelling}, are recognized for their effectiveness in providing more accurate food security risk assessments while addressing model uncertainty.  While the majority of such research focuses on assessing global food security risks, national food security is a pivotal concern in sustainability studies, with population dynamics playing a crucial role in this context \cite{garnett2013,premanandh2011}. To this respect, other studies that explore food risks at regional or national levels are \cite{chen2021disclosing, mainuddin2015national}. In this study, we implement a scenario-based approach for assessing food security risk at a national level, utilizing a diverse array of socio-economic and climate scenarios within a probabilistic framework. This method, where each scenario is treated as a model capable of generating multiple trajectories, is a novel approach in the field. We have adapted the standard framework of the Shared Socioeconomic Pathways to align with this probabilistic setting. 

\subsection{The main probabilistic population model}\label{RAFTERY-POP} 

The dynamics of population growth and the evolution of age structures are fundamental in driving numerous socioeconomic indices, including economic growth, production capacities, environmental challenges, and the demand for food and water. Consequently, demographic analysis is an essential starting point for any socioeconomic modeling study. In this section, we present some key findings related to the development of scenarios for future population growth. The leading model in this domain is the probabilistic approach proposed by Raftery and colleagues \cite{raftery2012}, which addresses the inherent uncertainties in demographic forecasting. This model employs a Bayesian hierarchical approach (referenced in \cite{congdon2010, gelman1995}) to account for the uncertainties and variability affecting future population projections.
The model is based on  the natural evolution of the population phenomenon as characterized by the standard (deterministic) model employed by the United Nations (UN), 
\begin{equation}\label{pop-model}
P_{c,t} - P_{c,t-1} = B_{c,t} - D_{c,t} + M_{c,t}
\end{equation}
where $P_{c,t}$ denotes the population of country $c$ at time $t$ (corresponding either to a single year or a 5-year period), $B_{c,t}$ stands for the number of births (which depends on the total fertility rate of the country), $D_{c,t}$ denotes the number of deaths (which depends on the life expectancy) and $M_{c,t}$ measures the net international migration. Uncertainty is introduced to the population model since its main components (fecundity, mortality, migration) are subject to random factors that cannot be sufficiently modelled. The Bayesian approach proposed in \cite{raftery2012} captures uncertainty on each one of the major components of population through the construction/introduction of distinct hierarchical models for important components such as fertility, mortality and migration\footnote{\scriptsize Note that only international migration is considered and not internal migration phenomena like urbanization}, and then propagating uncertainty to the output of model \eqref{pop-model}, i.e. providing probabilistic estimates either for $P_{c,t}$ or its breakdown into age groups and sex at various future times $t$. Clearly, uncertainty becomes higher as the time progresses. Based on an extensive database of past world population data (recorded population pyramids, fertility rates, mortality rates, etc), the fundamental law \eqref{pop-model}, and the principles of Bayesian statistics, the probabilistic features of the uncertainty factors driving the population fluctuations are recovered. Then, using this information, the fundamental law \eqref{pop-model} is iterated forward and used to obtain estimates for the future evolution of the quantities of interest. The estimates incorporate in a dynamically consistent fashion the effects of ambiguity as documented at least from the past data, and thus provide uncertainty consistent predictions for the future. 

One of the key features of the model is that it allows for quantities related to population projections to be random variables, characterized by a probability distribution, rather than point estimates. In particular, instead of producing a point estimate for a population related quantity  $X(t)$  at time $t$ ($X$ can represent for example population for a particular age group or sex, or quantities such as fertility, life expectancy, etc.), the model treats $X(t)$ as a random variable and produces (dynamically) a set of possible realizations $\{ X^{(j)}(t) \,\, : \,\, j = 1, \ldots, n\}$, which are approximations for the probability distribution of $X(t)$, based on possible outcomes of the uncertainty factors driving the phenomenon. Using this probability distribution, one characterizes the quantity $X(t)$ with quantities carrying more information than a mere point estimate, for example its percentiles at certain confidence levels or conditional means. These different realizations 
$$\Pi:=\{ X^{j}(t) \,\, : \,\,  t=T_0, \ldots, T, \,\, j=1,\ldots, n \},$$
where $T_0 < T$ are two selected time horizons, will be referred to as trajectories with $\{ X^{j}(t) \,\, : \,\, t=T_0, \ldots, T\}$ for fixed $j$ representing a particular realization (i.e. a particular possible path) for the evolution of population parameter in the future time interval $[T_0, T]$. Clearly, only one of the above paths in $\Pi$, if any, will materialize. However, the set of paths $\Pi$ provides us with important information concerning the probability of occurrence of paths with certain characteristics and allows for prediction of future population trends as well as the formulation of scenarios concerning these trends.

Some information on the structure of the probabilistic population model must be introduced here in order to make the SSP scenario generation procedure described in Section \ref{POP-SCENARIOS} more clear. In particular, \cite{raftery2012} rely on model \eqref{pop-model}, however treating the $B_{c}(t)$ and $D_{c}(t)$ components separately, according to the probabilistic modeling approach mentioned above. First, a hierarchical model is constructed for the Total Fertility Rate (TFR) component which provides projections for the fertility rates distribution at the country level and then  for the number of births distribution according to the approach presented in \cite{alkema2011}. Then, this information is used to an hierarchical model for the Life Expectancy (e0) component according to the approach presented in \cite{raftery2013}, which is then used to provide projections for life expectancy distributions of females and males per country at the various age-groups as well as to provide the mortality rates distribution on each age-group by gender. In particular, the life expectancy ($e_0$) and the total fertility rate (TFR) components of the model are captured by the parametric equations
	\begin{eqnarray}\label{AAAA}
		\begin{array}{ll}
			e^f_{0,c,t+1} = e^f_{0,c,t} + g_1( e^f_{0,c,t} \mid {\bm \theta}^{c}_{1} ) + \eta^{e_0}_{c,t+1}, & \eta^{e_0}_{c,t+1} \sim N\left(0, \varphi_{1}( e_{0,c,t}^{f}) \right)\\
			TFR_{c,t+1} = TFR_{c,t+1} + g_2( TFR_{c,t} \mid {\bm \theta}^{c}_{2} ) + \eta^{TFR}_{c,t+1}, & \eta^{TFR}_{c,t+1} \sim N\left( 0, \varphi_2(TFR_{c,t}) \right)\\
		\end{array}
	\end{eqnarray}
where the functions $g_1(\cdot), g_2(\cdot)$ determine separate double logistic-type growth models with respect to $e_0^f$ and $TFR$ expanding the UN modelling approach to a probabilistic setting (please see \cite{alkema2011,raftery2013} for details), respectively, while the functions $\varphi_1(\cdot), \varphi_2(\cdot)$ determine the variance terms concerning the residuals of each model. Note, that  in system \eqref{AAAA} the life expectancy model concerns only the females, while the life expectancy for the males ($e_0^m$) is obtained by building a gap model with respect to the $e_0^f$ term, according to the approach described in \cite{raftery2013}. Moreover, the whole modelling approach estimates a set of country-specific parameters (${\bm \theta}^{(c)} := ({\bm \theta}_1^{(c)}, {\bm \theta}_2^{(c)})'$) referring to the special characteristics of each country (as displayed by the available data) concerning life expectancy and total fertility rate, while each set ${\bm \theta}^{(c)}$ comes as a sample of a world distribution subject to some world-specific parameters (as obtained from the whole training dataset for the population model). In this perspective, the noise introduced in the population projections is carefully parameterized by the empirical evidence both on the total available dataset (world data) and on the characteristics displayed on a local level (country-specific attributes). 

\begin{figure}[ht!]
	\centering
	\includegraphics[width=4.2in]{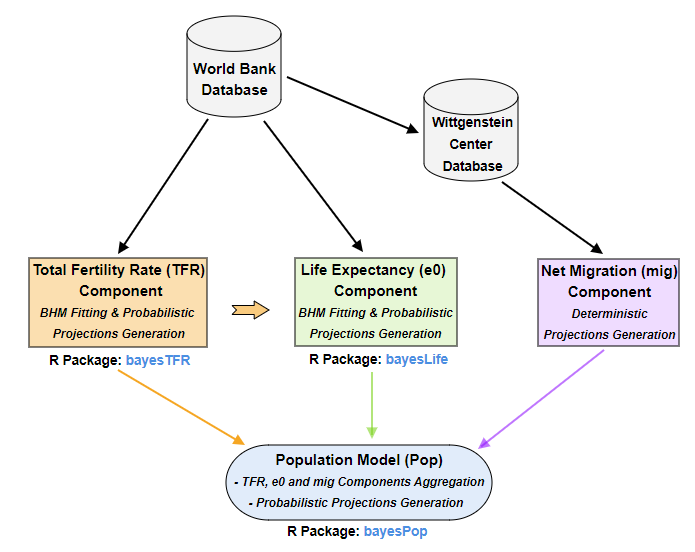}
	\caption{The procedure followed for generating probabilistic population projections illustrating the data providers, the separate model components and the respective R packages that are used.}\label{fig-1}
\end{figure} 

Concerning the third component that contributes to the population model, the net migration (MIG) term (at country level), projections for the future states are collected by the UN and other databases (see e.g. Wittgenstein Centre Database\footnote{\scriptsize \url{http://www.wittgensteincentre.org}}) and then incorporated to the main population model \eqref{pop-model}. Note that although there are similar hierarchical modeling approaches for migration in the literature (e.g. \cite{azose2015}) as the ones discussed above for the other two components, the lack or insufficiency of migration data for all the countries in our area of interest makes the implementation of this model infeasible at present, so we resort to the simple modelling framework mentioned in the previous sentence. Finally, all the above components are combined and aggregated in the general population model \eqref{pop-model} to provide the future population projections in terms of trajectories (possible scenarios) or distributions if conditioned to certain time instants. All the above modeling task is implemented in the statistical software R through the related package {\bf bayesPop} described in detail in  \cite{sevcikova2016}. The roadmap of the whole modeling task is illustrated in Figure \eqref{fig-1}.
 
\subsection{Shared Socioeconomic Pathways,  Population and GDP Projections}\label{POP-SCENARIOS}    

The practice of developing scenarios for future events has become integral in environmental economics, serving as a key tool for analyzing potential outcomes. A significant category of these scenarios is the Shared Socioeconomic Pathways (SSPs) \cite{oneil2014}, which outline plausible developments for critical socioeconomic drivers such as population growth and fertility in various regions of the world. These SSPs categorize potential global futures into five qualitative narratives—rapid, medium, stalled, inequality, and development—based on specific demographic characteristics like fertility, life expectancy (or mortality), migration, and, although not considered in our method, education.

These SSP narratives, presented qualitatively in Table \ref{tab-1.1}, require translation into quantitative terms for effective incorporation into quantitative models. For example, the SSP1 scenario describes the evolution of life expectancy in high-fertility countries as `high'. This leads to the question: What quantitatively constitutes `high' life expectancy in a specific country? How can we assign a definitive figure to this qualitative descriptor with an associated probability?' A pivotal aspect of our work is this translation of SSP narratives into a quantitative, probabilistic framework. This adaptation enables us to provide population projections that are quantified in terms of probability distributions or sample paths for the relevant demographics under each SSP pathway.

\begin{table}[ht!]\scriptsize
	\centering
	\begin{tabular}{l|cccc}
		\hline\hline
		\bf Socio-Economic Scenario &  \bf Country    & \bf Fertility & \bf Life       &  \bf Migration\\
		&  \bf grouping &                & \bf expectancy &                \\
		\hline
		& HiFert         & Low        & High & Medium\\
		{\bf SSP1:} \it Sustainability  & LoFert        & Low        & High & Medium\\
		& Rich-OECD  & Medium  & High & Medium\\
		\hline
		& HiFert         & Medium   & Medium & Medium\\
		{\bf SSP2:} \it Middle of the road        & LoFert        & Medium   & Medium & Medium\\
		& Rich-OECD  & Medium  & Medium & Medium\\
		\hline
		& HiFert         & High   & Low  & Low\\
		{\bf SSP3:} \it Fragmentation & LoFert        & High   & Low  & Low\\
		& Rich-OECD  & Low   & Low  & Low\\  
		\hline
		& HiFert         & High   & Low       & Medium\\
		{\bf SSP4:} {\it Inequality}                & LoFert         & Low   & Medium  & Medium\\
		& Rich-OECD  & Low   & Medium  & Medium\\  
		\hline
		& HiFert         & Low   & High  & High\\
		{\bf SSP5:} \it Conventional development & LoFert         & Low   & High  & High\\
		& Rich-OECD  & High   & High  & High\\                                                          
		\hline\hline                           
	\end{tabular}
	\caption{The Shared Socioeconomic Pathways (SSP) definitions}\label{tab-1.1}
\end{table}

The SSP scenarios are tailored to each country, classified as either a high fertility country (HiFert), a low fertility country (LoFert), or a Rich-OECD country, with specific SSP characteristics based on these groupings \cite{lutz2014, lutz2018}, as detailed in Table \ref{tab-1.1}. A significant issue with defining scenarios in this manner is the challenge in universally applying qualitative descriptors like low, medium, and high to different quantities of interest. This precision is crucial in quantitative modeling, which forms the foundation of policy-making.

Consider, for instance, the task of quantifying the 'low fertility' sub-scenario for both LoFert and HiFert countries. The UN's fertility modeling uses a threshold value of 2.1, interpreting values below this as indicative of a declining population and values above it as a growing population. However, when generating population scenarios for a relatively short term (say, 20-30 years), this fixed threshold may not accommodate all possible Total Fertility Rate (TFR) sub-scenarios due to varying data dynamics. For a LoFert country, a 'high TFR' case might be unrealistic, just as a 'low TFR' case might be improbable for a HiFert country. This rigid application of high-low thresholds can lead to a problematic situation where policymakers are unable to create comprehensive SSP scenarios for all countries across all time horizons, posing significant challenges in environmental modeling.

However, as we demonstrate in this section, adopting a probabilistic approach to modeling (like the population model described in Section \ref{RAFTERY-POP}) can effectively integrate with the qualitative SSP scenarios to transform them into well-defined quantitative tools. This approach offers a realistic and concrete framework for scenario building, where the various sub-scenarios (Low, Medium, and High) are endogenously and consistently determined by the system's evolving dynamics, rather than relying on rigid, deterministic cut-offs. This methodology enables more flexible and accurate scenario construction, crucial for effective environmental modeling and policy planning.
The quantification of SSP scenarios, as an alternative to deterministic cut-offs, can be effectively achieved through the probabilistic population model discussed in Section \ref{RAFTERY-POP}. This model allows for the generation of sample trajectories representing possible evolutions over time for fertility and life expectancy in each country of interest. By analyzing these samples, we can trace the evolution of the probability distribution of these demographic components over time. Subsequently, specific quantiles of these distributions at designated time points are utilized to establish the quantitative thresholds that define the {\it Low}, {\it Medium}, and {\it High} sub-scenarios, as referenced in Table \ref{tab-1.1}. For example, to determine the range for the three TFR sub-scenarios (Low, Medium, High) for a country $c$ by a certain year $T$ (e.g., $T=2050$), we use the distribution of the projected trajectories at $T$. These projections are divided into three equal segments based on the 33\% and 66\% quantiles. These quantile values then act as the boundaries differentiating the trajectory samples into the respective sub-scenarios. A trajectory is classified as {\it high} if, in the year 2050, its corresponding measure falls within the top 33\% of the empirical distribution. Specifically, for a trajectory $j$, it would be categorized as 'high' if $TFR_{c,2050,j} \geq q_{TFR,c,2050}(0.66)$, where $q_{TFR,c,2050}(\cdot)$ is the quantile function derived from the sample for TFR in 2050 for country $c$. The allocation of trajectories to the other two sub-scenarios, {\it Low} and {\it Medium}, follows a similar methodology.

\begin{table}[ht!]\small 
	\centering
	\begin{tabular}{c|c|cc}
		\hline\hline
		{\bf Population Driver} & {\bf Sub-Scenario} & \multicolumn{2}{c}{\bf Threshold Value}\\
		& & {\bf Lower} & {\bf Upper}\\
		\hline
		&Low         & & $ q_{TFR,c,t=T}(0.33)$\\
		Total Fertility Rate (TFR) & Medium & $q_{TFR,c,t=T}(0.33)$ & $ q_{TFR,c,t=T}(0.66)$\\
		& High       & $  q_{TFR,c,t=T}(0.66)$ & \\
		\hline
				&Low         & & $q_{e_0,c, t=T}(0.33)$\\
		Life Expectancy ($e_0$)  &Medium & $ q_{e_0,c,t=T}(0.33)$ & $ q_{e_0,c,t=T}(0.66)$\\
		&High       & $q_{e_0,c,t=T}(0.66)$ &\\
		\hline
					& Low & \multicolumn{2}{c}{\emph{As specified in} \cite{lutz2018} \emph{(deterministic)} }\\ 
		Net Migration (MIG)     & Medium & \multicolumn{2}{c}{\emph{As specified in} \cite{lutz2018} \emph{(deterministic)} }\\
					& High       & \multicolumn{2}{c}{\emph{ As specified in} \cite{lutz2018} \emph{(deterministic)} }\\
		\hline\hline
	\end{tabular}
	\caption{Population drivers sub-scenarios definitions based on the generated sample of trajectories}\label{tab-001}
\end{table} 

The quantification of SSP scenarios using our methodology offers a notable advantage over traditional approaches, such as the UN methodology. In our approach, the scenario levels (i.e. sub-scenarios) are not preassigned; instead, they are endogenously determined by the historical data dynamics captured within the probabilistic population model. This model, inherently Bayesian, is calibrated using an extensive global database of past population data, which embeds significant historical insights into the population phenomenon\footnote{The Bayesian model's reliance on a vast global database infuses it with rich historical context, making it highly informative.}. By applying this methodology to each trajectory in the sample for each country, we create three distinct sub-samples for both fertility rates and life expectancy. Each sub-sample corresponds to different potential realizations of the low, medium, and high intensity levels (sub-scenarios), providing critical statistical information like moments and variability within these scenarios. This approach offers a substantial advantage in the scenario-building process. It allows for the generation of more robust scenarios and the calculation of conditional expected values for other relevant quantities, based on the fundamental demographics modeled by these trajectories. A potential concern might be the representativeness of the Bayesian model's generated sample of trajectories for key population drivers (TFR, $e_0$) and its capacity to encompass all plausible future states. However, given the model's reliance on observed data from previous periods and its consideration of interconnections among countries and regions globally, it is reasonable to expect that any realistic case, reflective of the data available at the time of projection, can be simulated. Using a sufficiently large number of simulated trajectories, such as 100,000, should ensure reliability in the results. The specific country-level discrimination rules for each population driver are detailed in Table \ref{tab-001}. It is important to note that migration levels could also be determined in a similar probabilistic manner (as suggested by \cite{azose2015, azose2016}). However, for the sake of simplicity and due to the minor role of migration in our application, we utilize pointwise net migration projections under each SSP scenario, as provided by the Wittgenstein Center database. 

\begin{figure}[ht!]
	\centering
	\includegraphics[width=5in]{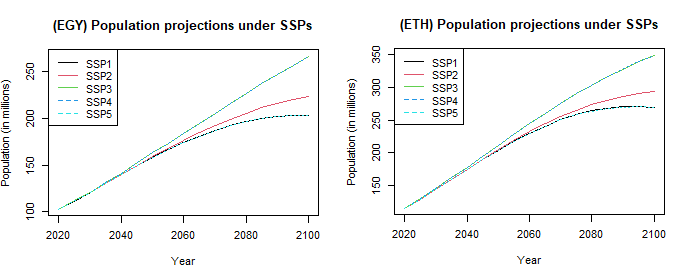}
	\caption{Median probabilistic population projections under each SSP scenario for Egypt and Ethiopia.}\label{fig-pop}
\end{figure}

One can apply this methodology to a country of interest, thereby creating an extensive database of potential future population outcomes as derived from the population model \eqref{pop-model}. This database aligns with the various SSP scenarios and their respective sub-scenarios, focused on key population parameters. For instance, in a country categorized under the HiFert group (as per Table \ref{tab-1.1}), the SSP2 scenario would encompass the TFR trajectories that fall within the Medium TFR sub-scenario, the e0 trajectories within the Medium e0 sub-scenario, and the net migration projections under the medium sub-scenario. These individual components – TFR, e0, and net migration – are integrated within the population model \eqref{pop-model} to construct the comprehensive population trajectories that constitute the SSP2 scenario for this specific country. This probabilistic process, enables the calculation of various statistical measures, such as quantiles and moments for different population dimensions specific to each country. An example of the application of this methodology is illustrated in Figure \ref{fig-pop}, which displays the median population projections for Egypt (EGY) and Ethiopia (ETH) under each SSP scenario. Additionally, Figure \ref{EGY-ETH-unc} in the Appendix \ref{App-A} presents the relative growth rates of these populations, along with the 90\% uncertainty zones, for each scenario, while the instantaneous projected total population distributions at year 2050 for Egypt and Ethiopia under all SSP scenarios are illustrated in Appendix \ref{App-B}.

Building scenarios representing different pathways for the population and its age structure evolution, offers a vehicle for the estimation of key economic drivers, like labour force, gross domestic product (GDP) and others (see e.g. \cite{aksoy2019demographic}). In this paper such economic drivers scenarios (i.e. employing the various SSPs) are obtained using the global dynamic economic model MaGE \cite{foure2013, foure2021}. This model was developed by J. Four\'e, A. B\'enassy-Qu\'er\'e and L. Fontagn\'e and takes into account factors like aging populations, changes in fertility rates, and migration, and how these demographic shifts impact economic growth. \footnote{\scriptsize It is freely available from CEPII's website\footnote{\scriptsize \url{http://www.cepii.fr/cepii/en/bdd_modele/presentation.asp?id=13}}} MaGE assumes that the world consists of economies of individual countries with each country $c$  characterized at time $t$ by a three-factor CES (Constant Elasticity of Substitution) production function with the capital and labour contributions modelled by the Cobb-Douglas parametric form 
\begin{eqnarray*}
	I_{c,t}=\left\{ (A_{c,t} \,\, K_{c,t}^{\alpha} \,\, L_{c,t}^{1-\alpha} )^{\frac{\sigma-1}{\sigma}} + (B_{c,t} E_{c,t})^{\frac{\sigma-1}{\sigma}} \right\}^{\frac{\sigma}{\sigma-1}}, \,\,\, 
	0 < \alpha <1,  \,\, 0 < \sigma <1
\end{eqnarray*}
where $I$ denotes the GDP, $K$ the capital, $L$ the labour force\footnote{\scriptsize child labour is not modelled by MaGE}, $E$ the energy consumption with $c$ denoting the country and $t$ corresponding either to 1-year periods or 5-years periods. The elasticity parameters are assumed to be the same for all countries ($\alpha=0.31$, $\sigma =0.136$) and constant in time, while the parameters $A$ (Total Factor Productivity or TFP) and $B$ (Energy Productivity or Energy Efficiency\footnote{\scriptsize where energy efficiency is defined as an analogous quantity to the ratio between energy consumption and GDP}) are assumed to be country specific and temporally varying. The model depends on population primarily through labor force and secondarily through the life cycle savings modelling which is introduced in the modelling of investment. MaGE allows the user to run generalized scenarios concerning the future states of the world economy, compatible with the SSP scenarios as far as population is concerned and produce relevant projections for each SSP scenario. In particular the model included in MaGE allows for the integration of demographics (including education) and economics to allow for predictions concerning population growth and economic quantities such as economic growth rates, GDP, energy consumption, etc.

\subsection{A Combined Framework with Climate Scenarios}

 The impact of climate change on food security risk assessments has been a focal point in much of the relevant literature, as evidenced by studies like \cite{schmidhuber2007global}. Key discussions on the necessity of determining effective measures and actions to address food security challenges arising from climate change and extreme climate events are presented in \cite{campbell2016reducing} and \cite{hasegawa2021extreme}. However, \cite{hasegawa2018risk} suggest that the adverse effects of stringent climate mitigation policies could outweigh the direct impacts of climate change on global hunger and food consumption. Understanding how these climatic factors interact with population dynamics is crucial for anticipating future food security challenges. In the context of food security issues, this means that it  important to consider critical environmental drivers such as temperature, precipitation, and water stress. These factors directly impact agricultural operations and, consequently, the overall capacity of the food system. To this end, the Representative Concentration Pathways (RCPs) \cite{van2011representative} provide essential scenarios. These scenarios project the increase in the Earth's mean temperature by the year 2100, factoring in the implementation of various environmental policies. The RCPs are characterized by the degree of temperature increase and the extent of mitigation measures required in each scenario. For detailed specifications of each RCP scenario, please refer to Table \ref{tab-1.2} in the Appendix \ref{App-A}.

\begin{table}[ht!]\small
\centering
\begin{tabular}{lccc}
\hline\hline
\bf Scenario & \bf SSP scenario & \bf RCP scenario & \bf Environmental Interpretation \\
\hline
SSP1-1.9  & SSP1 & RCP 1.9 & Most optimistic scenario\\
SSP1-2.6 & SSP1 & RCP 2.6  & Second-best scenario\\
SSP2-4.5  & SSP2 & RCP 4.5 & Middle of the road scenario\\
SSP3-7.0  & SSP3 & RCP 7.0 & Baseline of worst-case scenarios\\
SSP4-6.0  & SSP4 & RCP 6.0 & Best-case of the worst-case scenarios\\
SSP5-8.5  & SSP5 & RCP 8.5 & Worst-case scenario\\
\hline\hline
\end{tabular}
\caption{List of the combined SSP-RCP scenarios investigated in this work}\label{tab-1.3}
\end{table}

The current trend in developing realistic future world scenarios involves the integration of the Shared Socioeconomic Pathways (SSP) with the Representative Concentration Pathways (RCP). Initially, it might seem straightforward to conceive 30 distinct SSP-RCP scenarios, considering the ones outlined in Table \ref{tab-1.1} and the six standard RCP scenarios. However, the interaction between SSP and RCP scenarios is more complex, as they are interrelated despite focusing on different target quantities. Significant insights into these combined scenarios have been provided by the Coupled Model Intercomparison Project (CMIP)\footnote{\scriptsize \url{https://www.wcrp-climate.org/wgcm-cmip/wgcm-cmip6}}, which integrates diverse environmental models and utilizes extensive databases. The well-tested and available combined SSP-RCP scenarios, which include six distinct scenarios reflecting varying degrees of optimism about temperature change, are detailed in Table \ref{tab-1.3} \cite{meins2020}.

In line with our objective to assess food security risks within a framework that aligns with both socioeconomic and climate pathways, we conduct our estimations based on these six combined scenarios. This involves developing appropriate models for food capacity, tailored to each scenario, as discussed in Section \ref{model}).

\section{Towards a National Food Security Risk Index}\label{model}

The World Food Summit \cite{summit1996rome} defines food security as \emph{"when all people, at all times, have physical and economic access to sufficient, safe and nutritious food that meets their dietary needs and food preferences for an active and healthy life".} According to this definition, food security encompasses four distinct dimensions\footnote{\scriptsize \url{https://www.worldbank.org/en/topic/agriculture/brief/food-security-update/what-is-food-security}}: (a) the physical availability of food, (b) the economic and physical access to food, (c) the utilization of food, and (d) the stability and sustainability of the other three dimensions over time. 

In this context, we consider two primary metrics as the main components of food security risk: (a) the population's minimum food requirements at a specific time point (focusing on caloric content rather than recommended dietary composition), and (b) the capacity of the national food system at the same time, quantified as the total caloric content of food available to the population as a result of the system's functioning. It is important to note that our approach does not directly address finer aspects such as dietary needs, food habits, or poverty-related access issues at the household- or individual-level.\footnote{\scriptsize Examples include Australia, Canada, the UK, and the USA, where high national food security contrasts with notable household-level food insecurity (see e.g. \cite{loopstra2018interventions})}. Nevertheless, the framework we propose is adaptable and can be extended to include these and other factors, should they be necessary for a more nuanced analysis. Henceforth, we use the term 'food requirements' to refer specifically to caloric needs or intake.

Forecasting future caloric shortages is crucial for policymakers, enabling them to implement proactive strategies such as restructuring food production, optimizing land use, or planning imports. Developing effective policies requires a deep understanding of the dynamics influencing significant shifts in the food system. In this section, we introduce a streamlined statistical model that utilizes minimal and publicly available data to generate country-level, probabilistic projections of minimum caloric requirements and food system capacity, thereby quantifying food security risk. The model’s primary driver, population growth, is approached probabilistically, building upon the detailed model by \cite{raftery2013} and its adaptations discussed in Section \ref{POP-MODELS}. While socioeconomic factors (like GDP and labor), natural resources (such as water resources and cropland area), and climate variables (temperature, precipitation) are integral to the model, they are not treated probabilistically. This limitation stems from constraints in computational resources, data availability, or the absence of suitable models, which would otherwise make the model excessively complex and challenging to interpret. Therefore, probabilistic scenarios are predominantly applied to population growth, the main driver. Expanding the probabilistic treatment to other drivers remains an objective for future research, contingent upon the availability of resources and appropriate models.

Our modelling approach is designed to create a food security risk index that is both reliable and robust against model uncertainties. This index assesses a country's capacity to meet its population's minimum caloric needs through the current function of its food system. The index's construction is based on easily accessible, reliable data that span a significant historical period for numerous countries. It is tailored to project into the future under combined SSP-RCP scenarios, taking into account each scenario’s impact on the contributing factors to the index. It is important to note that in this version of the index, dietary patterns, food habits, and food affordability factors are not incorporated, with a focus instead on the core elements of food security risk assessment.

\subsection{Estimating minimum caloric requirements}\label{NEED-FOR-FOOD}

Minimum food requirements for subsistence, which are distinct from the caloric intake necessary for nutritional security\footnote{\scriptsize It's important to note that the bare minimum caloric intake for survival does not align with the calorie requirements ensuring nutritional security}, are intrinsically linked to population size. These requirements are also influenced by the detailed age structure and activity levels of each age group. Humans have a fundamental requirement for a specific range of daily caloric intake, reflecting the inelastic nature of the need for food for subsistence This requirement varies across different age groups, genders, and lifestyles, notably influenced by levels of physical activity. Typically, the daily caloric intake necessary for an individual can range from approximately 1000 to 3200 calories, contingent upon these factors. In the provided Tables \ref{tab-male-intake} and \ref{tab-female-intake}, we detail these caloric requirements. These tables reflect recommendations from the United States Department of Health and Human Services and Department of Agriculture (HHS/USDA)\footnote{\scriptsize United States Department of Health and Human Services and Department of Agriculture \url{https://www.dietaryguidelines.gov/sites/default/files/2020-12/Dietary_Guidelines_for_Americans_2020-2025.pdf}}, categorizing the necessary daily calorie intake for both male and female populations, segmented by age group and level of physical activity. Given the age structure of the male and female population (i.e. population pyramids per gender) for a country $c$, we may then obtain an estimate for the total daily recommended calories intake at year $t$, through the quantity
\begin{eqnarray}\label{TOTAL-CALORY-INTAKE}
	C^{R}_{c,t}=\sum_{a} R_{a}^{f} P_{a,c,t}^{f} + \sum_{a} R_{a}^{m} P_{a,c,t}^{m}
\end{eqnarray}
where $a$ corresponds to the age groups mentioned in Tables \ref{tab-male-intake} and \ref{tab-female-intake}, $P_{a,c,t}^{m}, P_{a,c,t}^{f}$ denote the total male and female population for the respective age groups and $R_{a}^{m}, R_{a}^{f}$ represent the calories requirements given in Tables \ref{tab-male-intake} and \ref{tab-female-intake}. This estimate varies, depending on the activity level distribution of the population, however, one may obtain a lower bound for this quantity using the values for $R_{a}^{m},R_{a}^{f}$ for non-active individuals or an upper bound using these values for the very active individuals. In this perspective, the quantity $C^{R}_{c,t}$ acts as a proxy for the total daily calorie needs (in terms of a lower or an upper estimate). Clearly, the current estimate may deviate from the actual daily minimum calorie requirements on account of malnutrition issues related to diet patterns, poverty or unequal income distribution, etc. However, as actual data for total calorie needs are not available, we consider $C^{R}_{c,t}$ as a reasonable proxy.

Based on the above methodology, it becomes obvious that neither diet patterns or food product substitutions (e.g. when pasta is unavailable, replacing with rice, etc), or access to food are taken into account in this approach. The resulting estimate $C^R_{c,t}$ stated in \eqref{TOTAL-CALORY-INTAKE} does not necessarily coincide either with the minimum nutritional needs or the food consumption of the country under study. Nutritional needs are subject to dietary patterns, available crops or food types. The estimation of minimum nutritional needs is a much more complex task than the estimation of the minimum calorie intake, since it relies on several aspects of food demand. Similarly, food consumption also involves aspects of food demand and supply (see e.g. \cite{hasegawa2015scenarios, valin2014future, van2021meta}). However, the estimate $C_{c,t}^R$ provided here, depends directly only on population structure of the country under study and does not take into account other factors, in an attempt to provide a rough, but as reliable as possible, estimate for the caloric requirements of the country's population for subsistence. Clearly, a more detailed model could be considered as a next step in this modelling component, employing more detailed data concerning the nutrition patterns and habits, and the effects of active policies for increasing food affordability of the general population.

Utilizing the outlined methodology, it is important to acknowledge that this approach does not factor in dietary patterns, food product substitutions (such as replacing pasta with rice when the former is unavailable), or accessibility to food. The estimated total caloric requirement $C^R_{c,t}$, as detailed in \eqref{TOTAL-CALORY-INTAKE}, is not synonymous with either the minimum nutritional needs or the actual food consumption of the country being studied. Nutritional requirements are influenced by factors like dietary habits, availability of various crops, and types of food. Estimating minimum nutritional needs is inherently more complex than calculating minimum calorie intake, as it entails a broader array of food consumption aspects,  Similarly, such as demand and supply elements in the food system, as discussed in  \cite{hasegawa2015scenarios, valin2014future, van2021meta}. However, the estimate $C_{c,t}^R$ we provide is based solely on the population structure of the country in question, deliberately excluding other variables. This approach aims to offer a basic yet reliable estimation of the population's caloric requirements for subsistence. It's clear that a more comprehensive model, incorporating detailed data on nutrition patterns, eating habits, and the impact of policies aimed at improving food affordability for the general population, could be developed as a subsequent enhancement to this component of the modeling process.

\begin{table}[ht!]\scriptsize
\centering
\begin{tabular}{l|lll}
\hline\hline
\bf Age & \bf Not Active & \bf Somewhat Active & \bf Very Active  \\
\hline
2--3 years              & 1,000--1,200 calories  & 1,000--1,400 calories & 1,000--1,400 calories  \\
4--8 years              & 1,200--1,400 calories  & 1,400--1,600 calories & 1,600--2,000 calories  \\
9--13 years            & 1,600--2,000 calories  & 1,800--2,200 calories & 2,000--2,600 calories  \\
14--18 years          & 2,000--2,400 calories  & 2,400--2,800 calories & 2,800--3,200 calories  \\
19--30 years          & 2,400--2,600 calories  & 2,600--2,800 calories & 3,000 calories        \\
31--50 years          & 2,200--2,400 calories  & 2,400--2,600 calories & 2,800--3,000 calories  \\
51 years and older & 2,000--2,200 calories  & 2,200--2,400 calories & 2,400--2,800 calories\\
\hline\hline 
\end{tabular}
\caption{Calories Needed Each Day for Boys and Men  (Source: HHS/USDA Dietary Guidelines for Americans, 2010)}\label{tab-male-intake}

\begin{tabular}{l|lll}
	\hline\hline
	\bf Age & \bf Not Active & \bf Somewhat Active & \bf Very Active \\
	\hline
	2--3 years          & 1,000 calories       & 1,000--1,200 calories & 1,000--1,400 calories  \\
	4--8 years          & 1,200--1,400 calories & 1,400--1,600 calories & 1,400--1,800 calories  \\
	9--13 years         & 1,400--1,600 calories & 1,600--2,000 calories & 1,800--2,200 calories  \\
	14--18 years        & 1,800 calories       & 2,000~ calories      & 2,400~ calories        \\
	19--30 years        & 1,800--2,000 calories & 2,000--2,200 calories & 2,400~ calories        \\
	31--50 years        & 1,800 calories       & 2,000~ calories       & 2,200~ calories       \\
	51 years and older & 1,600 calories       & 1,800~ calories      & 2,000--2,200 calories \\
	\hline\hline
\end{tabular}
\caption{Calories Needed Each Day for Girls and Women (Source: HHS/USDA Dietary Guidelines for Americans, 2010)}\label{tab-female-intake}
\end{table}

The population model (see Section \ref{RAFTERY-POP})  provides accurate probabilistic predictions for the population pyramid, i.e., for the quantities $P_{a,f,c,t}, P_{a,c,m,t}$. Using this model, a number (let us say $M$) of different realizations for the population pyramid in terms of data batches of trajectories are obtained, concerning the evolution of both female and male population per age group over the time period $[T_0, T]$, i.e.
$$ P_{f}(\tau, \mathcal{J}):=\left\{ P_{a,c,f,t}^{(j)}\, :\, j\in \mathcal{J}, \,\, t\in\tau \right\}, \,\,  P_{m}(\tau, \mathcal{J}):=\left\{P_{a,c,m,t}^{(j)}\, :\, j\in \mathcal{J}, \,\, t\in\tau \right\}$$ 
where $\tau = T_0,...,T$ and $\mathcal{J}=\{1,2,...,M\}$. As already stated, the uncertainty effects are properly accounted for in these trajectories and in accordance to past data. Taking a slice of, e.g., $P_{f}(\tau, \mathcal{J})$ at a fixed time $t' \in [T_0, T]$, will provide the sample $P_{f}(t', \mathcal{J})$ consisting of $M$ possible realizations of the  random variable $P_{a,c,f,t'}$ which can provide useful information concerning its distribution (i.e. moments, quantiles, etc). In fact, the general trajectories can be classified according to various criteria that characterize the SSP scenarios (see Section \ref{POP-SCENARIOS}) so as to obtain subsets of the trajectories which are compatible with the various SSP scenarios. Using the trajectories for each scenario we may obtain conditional means or quantiles for the conditional distribution of the quantities $P_{a,c,f,t'}, P_{a,c,m,t'}$ per SSP scenario. This procedure allows us to have a detailed probabilistic scenario-based description of the possible evolution of future population related quantities.

After obtaining the trajectories and probabilistic scenarios for the population and its age structure, using the estimate \eqref{TOTAL-CALORY-INTAKE} for the minimum caloric requirements, we may generate similar probabilistic scenarios for $C^{R}_{c,t}$ for the future. To this aim, we have to use the generated data batches $P_{f}(\tau, \mathcal{J})$, $P_{m}(\tau, \mathcal{J})$, and feed them into  formula \eqref{TOTAL-CALORY-INTAKE} to generate trajectories $C(\tau, \mathcal{J}) := \{C^{R}_{c,t,j}\, : \, t\in \tau, \,\, j\in\mathcal{J} \}$ where $C^{R}_{c,t,j}$ denotes the calorie requirements for the country $c$ at time $t$ according to the population pyramid generated in the $j$-th trajectory of the sample. These data batches will be subsequently used to generate samples for projections of $C^{R}_{c}$ on various future dates $T_0 \leq t' \leq T$, and from those as described above, probabilistic information on this important quantity will be generated. Clearly, when this quantity is needed in the context of SSP scenarios, the relevant trajectories for the  population quantities corresponding to these scenarios will be employed to the generation of trajectories in \eqref{TOTAL-CALORY-INTAKE} for estimating the minimum caloric requirements.

\subsection{Modelling food system capacity}\label{tfs-mod}

Food security risk is influenced not only by minimum food requirements but also critically by the capacity of national food systems. A potential risk emerges when a country's food system capacity, defined in terms of available calories for human consumption, fails to meet its population's minimum caloric requirements $C^{R}_{c,t}$. The Food and Agriculture Organization (FAO) succinctly defines the concept of food self-sufficiency\footnote{\scriptsize FAO (1999) \url{https://www.fao.org/3/X3936E/X3936E00.htm}} as "\emph{the extent to which a country can satisfy its food needs from its own domestic production}". However, self-sufficiency is not a synonym of food security in the sense that a nation's strategy may not solely rely on domestic production. Global markets always function and play a crucial role in ensuring food availability, especially in scenarios where local production is insufficient or disrupted. Thus, the capacity of a food system is significantly influenced not only by domestic factors but also by the country's economic engagement with global markets. This includes the planning of economic sectors involved in food production and distribution, choice of crops, and crucially, the dynamics of imports and exports. Acknowledging this broader perspective is essential for a comprehensive understanding of national food security and its reliance on both domestic supply and international market interactions. Socioeconomic conditions, natural resources, and climate factors all play pivotal roles in shaping the capacity of these food systems. While other dimensions of food security, such as physical, social, and economic access to food, sustainability, and nutrition, are undeniably crucial \cite{Burlingame2014,garnett2013}, they are often subsequent considerations and at the same time policy areas for addressing issues stemming from inadequate food system capacity.

Our method for assessing food security risk is designed to complement existing models, which often focus on immediate, short-term factors \cite{who2021}. In line with Ericksen's food system framework \cite{Ericksen2008}, we identify two primary groups of long-term drivers influencing food system outcomes: global environmental change drivers, including variables such as land cover changes, climate variability, and water and nutrient availability; and socioeconomic drivers, encompassing factors like demographics, economic conditions, and technological progress. Additionally, the High Level Panel of Experts on Food Security and Nutrition expands on this classification, categorizing these drivers into six broader areas: biophysical and environmental; technology and innovation; economic and market; political and institutional; socio-cultural; and demographic \cite{hlpe2017}. This comprehensive approach enables a more holistic understanding of the multifaceted factors impacting food security over extended periods.

To evaluate the implications of SSP-RCP scenarios on food security, we utilize a broadly structured model that reflects the influences of key drivers on the capacity of the food system. Our focus is on how shifts in socioeconomic elements (like population dynamics, income levels, and labor force) and environmental factors (such as land and water resources, along with changes in precipitation and temperature) shape the food system over time. These variables play critical roles in food-related activities, including production, consumption, market dynamics, and trade. The model aligns with available or projectable data for these factors under each SSP-RCP scenario. However, the inclusion of additional economic drivers, for instance, land use patterns and the workforce engaged in food production, while valuable \cite{gaitan2019, hlpe2017, vanberkum2018}, faces practical limitations. Detailed data on these aspects are often not accessible, especially for non-OECD countries. Furthermore, forecasting these variables for future scenarios poses significant challenges, both in terms of the complexity of economic modeling and the computational resources required.
	
To effectively model the food system capacity, our approach integrates a balanced interplay of domestic supply, imports, exports, and inferred demand. The first component, domestic food production, is contingent upon a country's natural resources like land and water, the labor force in agriculture or livestock sectors, and the prevailing economic conditions. The second component, while not directly measuring food demand due to data limitations, uses domestic food supply, imports, and exports as proxies to approximate demand, underpinning the equilibrium state of the food system's capacity. The latter part of the model acknowledges that food system capacity observed is an equilibrium between the amount of food a country produces, the food it imports and exports, and the overall demand within the country. Historical data on food system capacity are sourced from the FAO database\footnote{\scriptsize FAO STAT Web Database: \url{https://www.fao.org/faostat/en/\#data/SCL}} (referred to as total food supply in the database).

In \ref{upper-lm}-\ref{lower-lm}, we represent that model in a two-layer structure. This structure is primarily a representational tool, designed to enhance clarity and organization in the presentation of the model. It is important to clarify that these layers do not imply any hierarchical or sequential order in the estimation process, nor do they introduce any additional complexity or assumptions into the modeling approach. In essence, the division into 'lower' and 'upper' layers is a conceptual framework used for better illustrating how different types of drivers—socioeconomic and climate-related—affect the food system capacity. The lower layer, focuses on socioeconomic drivers, and the upper layer, encompassing both socioeconomic and environmental factors. This distinction is made to aid in the understanding and explanation of the model, rather than to suggest a layered approach in the actual computation or estimation of food system capacity. All factors, regardless of the layer they are categorized in, are integral to the model and contribute collectively to the estimation of food system capacity and, by extension, to the assessment of food security risk. Under the aforementioned considerations, we set the two-layer model:
\begin{eqnarray}\label{upper-lm}
 	&& \mbox{\bf Upper Layer } \left\{ 
 	\begin{array}{ll}
 	Q^{FSC}_{c,t} &= f_1\left(t, Q^{Dom}_{c,t}, Q^{Exp}_{c,t}, Q^{Imp}_{c,t} \right)\\
 	Q^{Dom}_{c,t} & =  f_2(t, P_{c,t}, I_{c,t-1}, L^{Agr}_{c,t}, W_{c,t}, A_{c,t}, T_{c,t})\\
 	W_{c,t}   &= f_3\left(t, P_{c,t},  I_{c,t-1}, Q^{Dom}_{c,t}, A_{c,t}, T_{c,t}, Pr_{c,t} \right)\\
 	A_{c,t} & =  f_4(t,P_{c,t}, I_{c,t-1}, Q^{Dom}_{c,t-1} )
 	\end{array}
 	\right.\\
 	&& \mbox{\bf  Lower Layer } \left\{
 	\begin{array}{ll}\label{lower-lm}	     
 	Q^{Imp}_{c,t} &=  g_1(t,P_{c,t}, I_{c,t-1})\\
 	Q^{Exp}_{c,t} & =  g_2(t,P_{c,t}, I_{c,t-1})\\
 	L^{Agr}_{c,t} &=  g_3(t,P_{c,t}, I_{c,t-1}, L_{c,t})
 	\end{array}
	\right.
\end{eqnarray}
where the upper layer concerns the modelling of quantities:
\begin{itemize}
	\item $Q^{FSC}_{c}$: the food system capacity of country $c$ at time $t$,
	\item $Q_c^{Dom}$: the domestically produced food quantity,
	\item $W_{c}$: the level of water stress of country $c$ at time $t$, i.e. the freshwater withdrawal in percentage of the available freshwater resources (according to the SDG Indicator 6.4.2\footnote{\scriptsize \url{https://www.fao.org/publications/card/en/c/CA8358EN/}}),
	\item $A_c$: the land area occupied for agricultural activities,
\end{itemize}
while the lower layer concerns the modelling of the quantities:
\begin{itemize}
	\item $Q^{Exp}_c, Q^{Imp}_c$: the exported and imported food quantities, respectively, and
	\item $L_c^{Agr}$: the labour force occupied in the food production (agricultural) sector, 
	\item $P_{c,t}$: The population of country $c$ at time $t$
\end{itemize}
which are subsequently fed into the upper layer to produce food system capacity estimates. 

\begin{figure}[ht!]
\centering
\includegraphics[width=5in]{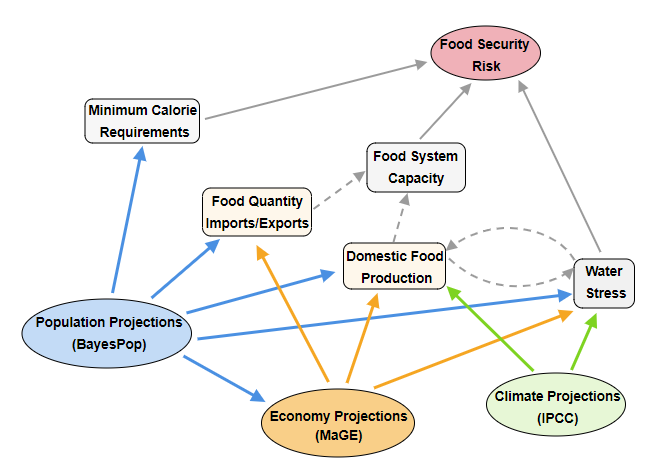}
\caption{The modelling approach concerning the food system capacity and minimum calorie requirements, illustrating inter-dependencies between the various model components and the main effects introduced by the various socioeconomic and climate drivers}\label{fig-fsri-model}
\end{figure}

In our model, the quantities in the lower layer are directly influenced by population ($P$), labor ($L$), and GDP ($I$), with their trajectories varying according to the specific SSP scenario being considered (SSP-dependent quantities). Conversely, in the upper layer, the quantities are not only affected by population ($P$) and GDP ($I$) but also by environmental variables like temperature ($T$) and precipitation ($Pr$), which are influenced by the chosen RCP scenarios. While the lower layer assumes no interdependencies among its various factors, the upper layer does account for interactions between the modeled quantities (a graphical representation of this modeling approach can be found in Figure \ref{fig-fsri-model}). Furthermore, additional assumptions or constraints on the relationships between these quantities can be implemented through suitable choices of functions $f_1, f_2, f_3$ for the lower layer and $g_1, g_2, ..., g_4$ for the upper layer. These might include, for example, specific evolutionary forms for the agricultural labor force considering technological changes, or policy or physical constraints on land and water resource usage. It is important to note that, although the case studies in this paper (refer to Section \ref{sec-4}) employ Cobb-Douglas type models, the model is flexible enough to accommodate other econometric approaches such as Translog models or CES production functions, depending on data availability.

The modeling framework outlined in \eqref{upper-lm}-\eqref{lower-lm} seeks to delineate the processes leading to the observed capacity of the food system, drawing on a range of socioeconomic and environmental drivers. Population plays a multifaceted role, influencing both labor force and income. Similarly, natural and environmental resources are key determinants of domestic food production. The food system's capacity emerges from the equilibrium between domestic food production and the interplay of food imports and exports. This balance reflects a country's participation in the global food market and its economic capability, influenced by GDP, to augment domestic production and satisfy demand.  In our approach, these factors are modeled at a regional scale, without incorporating global influences such as market interdependencies, which presents an avenue for future exploration. Additionally, GDP is introduced with a one-step delay in the model to capture its effect on food system activities like domestic production and trade, while simultaneously mitigating potential reverse causal impacts of these activities on GDP (\cite{zestos2002}). In the following section, we will discuss how these foundational assumptions are instrumental in enabling projections of food capacity that are vital for crafting a risk measure. This measure is designed not just to signal vulnerabilities but to also catalyze preemptive strategies that could fortify a nation's food security stance.

Our model operates under the implicit assumption that the food system will adapt gradually to changes in key drivers, as outlined by SSP-RCP scenarios, without experiencing abrupt structural shifts in activities such as demand patterns, production technology, crop usage, and food trade balance. Furthermore, it assumes that factors not explicitly modeled will remain up to time trend, captured by $t$. This approach, while simplifying the complexity of real-world dynamics, strategically identifies variables that policymakers have the capacity to influence. As detailed in the ensuing section, this modeling choice allows policymakers to utilize the food security risk index as a diagnostic tool. Should the index reveal areas of concern, the unmodeled drivers offer a set of adjustable parameters through which policymakers can enact changes to strengthen food security measures

\subsection{A food security risk index within and across socio-economic scenarios}\label{food-security-indices}

Concerning the task of quantifying food security risk, several indicators have been introduced in the literature so far. The United Nations has introduced the SDG Indicator 2.1.1, also known as the Prevalence of Undernourishment (PoU) indicator. This indicator is defined\footnote{\scriptsize \url{https://unstats.un.org/sdgs/metadata/?Text=&Goal=2&Target}} as `\emph{an estimate of the proportion of the population whose habitual food consumption fails to provide the necessary dietary energy levels for maintaining a normal active and healthy life, expressed as a percentage}'. However, the PoU indicator has faced criticism due to the extensive and detailed data requirements for its computation. It necessitates periodic household surveys, comprehensive information on food acquisition per household, dietary habits, and additional data on the total available food for human consumption to correct potential biases. Consequently, calculating this indicator is a complex task, requiring granular household-level information. Furthermore, forecasting future trends or immediate changes in this index poses significant challenges due to these data requirements.

Also, FAO's Food Insecurity Experience Scale (FIES) serves as an important tool in assessing food security, offering internationally comparable estimations of the extent of food access difficulties encountered by individuals and households.\footnote{\url{https://www.fao.org/policy-support/tools-and-publications/resources-details/en/c/1236494/}} It contributes to the monitoring of progress towards Sustainable Development Goal (SDG) Target 2.1, which is dedicated to ending hunger and ensuring universal access to food. The FIES derives its insights from direct interviews, quantifying the severity of food insecurity faced by people, and is instrumental in measuring the advances towards achieving SDG Target 2.1. This measure supplements the insights provided by the SDG Indicator 2.1.1, the Prevalence of Undernourishment (PoU) indicator, by capturing a more detailed picture of food access challenges.

Another prominent index is the Global Food Security Index (GFSI)\footnote{\scriptsize \url{https://impact.economist.com/sustainability/project/food-security-index/}}, which evaluates food affordability, availability, quality, safety, sustainability, and adaptation across 113 countries. The GFSI is a dynamic benchmark model, both quantitative and qualitative, comprised of 68 unique indicators that measure the drivers of food security in various countries, spanning both the developed and developing world. The complexity of estimating the GFSI surpasses that of SDG 2.1.1 due to its comprehensive nature, which aims to reflect the state of global food security. As a result, its utility lies primarily in presenting the current state of affairs rather than projecting future scenarios, given its intricate construction and wide-ranging scope.

Drawing upon the methodologies outlined in Sections \ref{NEED-FOR-FOOD} and \ref{tfs-mod}, which detail the calculations for minimum caloric requirements ($C^R$) and food system capacity ($Q^{FSC}$), we present a novel food security risk index in this section. This index is the result of an integrated approach, combining the aforementioned models to assess national-level food security risks. Our methodology is adept to evaluate future food security risks under various SSP-RCP scenarios, with the added benefit of requiring less detailed data inputs compared to established indices like the SDG 2.1.1 and FIES Indicators or the GFSI. A key strength of our proposed index is its robust treatment of uncertainty, particularly in the estimation of critical factors influencing food security. This includes addressing uncertainties in future population estimates, which are provided as probabilistic projections in our model. 

Building on the methodologies delineated in Sections \ref{NEED-FOR-FOOD} and \ref{tfs-mod}, which elaborate on the computation of minimum caloric requirements ($C^R$) and food system capacity ($Q^{FSC}$), this section introduces a new index for assessing food security risk. This index emerges from an amalgamation of the aforementioned models, facilitating the assessment of food security risks at a national level. Our model is adept at projecting future food security risks within diverse SSP-RCP scenarios, with the distinct advantage of minimizing the need for detailed data inputs, unlike more conventional indices such as the SDG 2.1.1 and FIES Indicators or the GFSI. A prominent feature of our proposed index is its rigorous approach to managing uncertainty, especially in the estimation of pivotal factors that impact food security. This involves incorporating uncertainties in future population figures, which are rendered through probabilistic projections in our model.

Our index is designed to be flexible and reliable, capable of assessing food security risks both within individual scenarios and across a range of different SSP-RCP scenarios. We take this opportunity here to clarify what we mean by within and across scenarios. For various reasons (e.g. the long-term horizons involved in the modelling of the phenomenon, or lack of sufficient data etc) it is not always possible for the decision maker to be aware of the exact scenario that will be materialised. Moreover, due to the long time horizon and the dynamic nature of the phenomenon, it may be that we start within the range of one scenario and in the course of time, as the phenomenon evolves, we enter within the range of a different scenario. This means that we should be able to make predictions concerning food security not only within a particular scenario (an SSP or an SSP-RCP scenario, i.e. referring to the ``within scenario'' case) but also taking in to account the possibility of possible multiple scenarios, or even transitions from a scenario to another. We will refer to the latter situation using the terminology ``across scenarios'', i.e. the situation where SSP or SSP-RCP mixed scenarios are considered, where each scenario in the set is considered as probable with a specified probability. The rationale behind this approach, is to provide a risk assessment tool capable of supporting the decision making process of policy makers, developing a working framework that allows to take into account the effects of important drivers and the uncertainty propagated to the food security risk assessment task. 

Keeping in mind the aforementioned considerations we define the new food security risk index. Given the minimum caloric requirements ($C^{R}$), the food system capacity ($Q^{FSC}$) and the water stress ($W$) of a country $c$ for a specific year $t$, the \emph{Food Security Risk Index (FSRI)} is defined as  
\begin{equation}\label{fsri}
	I^{FS}_{c,t} := \left[ \frac{1}{1+\gamma} \left( \frac{C^R_{c,t}}{ Q^{FSC}_{c,t}} \right)  + \frac{\gamma}{1+\gamma} W_{c,t} \right] \cdot 100\%
\end{equation}
where the sensitivity parameter $\gamma \in [0,\infty)$ denotes the relevant cost for country $c$ in acquiring extra water resources.

The proposed index expresses the percentage ratio between the minimum caloric requirements to the food system capacity at national (or regional) level, weighted by the level of water stress with respect to the sensitivity parameter $\gamma$. In the special case where the water stress risk is not taken into account (i.e. $\gamma \to 0$, assuming extremely low cost in acquiring extra water resources besides the available ones from the country, or a case of a country that will never face water stress issues), high capacity of the food system $Q^{FSC}$ comparing to $C^{R}$ leads to lower indicator values, while low $Q^{FSC}$ values comparing to $C^{R}$ should lead to increased food security risk. However, water stress risk in general significantly affects food security since most food production activities relies on water resources management and is of crucial importance, especially in areas where water resources are very limited, e.g. North Africa region. Clearly, the determination of the relevant sensitivity parameter $\gamma$ should be done with special care by the policy maker, taking into account the current and the future status of the under study area concerning the pressure on water resources and their necessity to the national or regional food production activities.

\begin{figure}[ht!]
	\centering
	\includegraphics[width=4.4in]{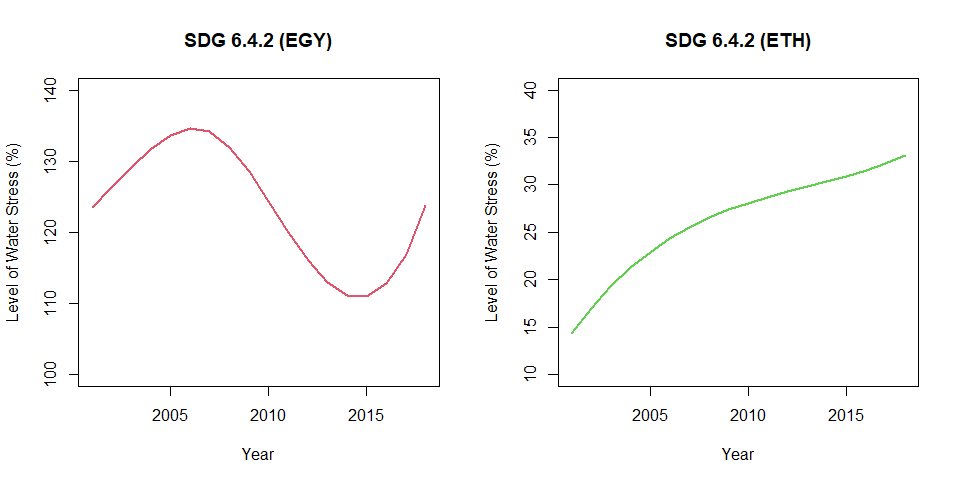}
	\caption{Level of water stress (SDG 6.4.2 Indicator) for Egypt and Ethiopia for the time period 2001-2018.}\label{fig-wsi-hist}
\end{figure}

Incorporating water stress explicitly into our food security risk index, despite its implicit consideration within the quantities $C^R$ and $Q^{FSC}$, enhances the model's robustness against potential misspecification issues. The modeling of water stress and food system capacity is inherently prone to errors due to unexpected socio-economic shifts (e.g., political unrest, wars), pandemics (like COVID-19), or climate change impacts (such as natural disasters), which can lead to transiently inaccurate estimates for food demand and supply patterns. These errors might result in unrealistic projections, such as an overestimated increase in food production that surpasses the limits of production mechanisms or natural resource availability, particularly water resources. The explicit inclusion of water stress in the model addresses this issue by directly considering the risks associated with water resource availability, especially in regions already experiencing high water stress\footnote{Countries with a water-stress indicator value above 40\% are considered as highly water-stressed}.  According to the World Resources Institute (WRI)\footnote{\url{https://www.wri.org/insights/highest-water-stressed-countries}}, water stress is classified into several levels: \emph{extremely high water stress} for values over 80\%, \emph{high water stress} between 40\%-80\%, \emph{medium-high water stress} from 20\%-40\%, \emph{low-medium water stress} between 10\%-20\%, and \emph{low water stress} below 10\%. In our case studies, Egypt and Ethiopia, each country exhibits distinct water resource pressures for the period 2001-2018 (Figure \ref{fig-wsi-hist}). Egypt falls into the \emph{extremely high water stress} category (WRI definition, $W>80\%$) with water stress levels fluctuating around 120-125\% in the last two decades. Ethiopia, initially a \emph{low-medium water stressed} country in the early 2000s, has shown an increasing trend, reaching approximately 35\% water stress in 2018, categorizing it as \emph{medium-high water stressed}. The divergent scenarios in these neighboring countries highlight the importance of including water stress data in the food security risk index to more accurately reflect the actual situation.

Our proposed index aggregates risks more broadly compared to detailed indices like SDG 2.1.1, necessitating a careful and meaningful selection of the parameter $\gamma$. It is important to note that our simpler index is particularly suited for future projections, a capability not as readily available in more complex indices. By calibrating the sensitivity parameter $\gamma$ to replicate features observed in these complex indices based on past data, we enhance confidence in the predictive potential of our index. According to the World Resources Institute (WRI) specifications for water stress intensity, our index can offer varied insights into the sensitivity to water resource pressures. A logical approach would be to adjust the value of $\gamma$ in relation to the deviation of the previously recorded water stress value from a defined safety threshold. Specifically, if the past water stress value falls below this threshold, the water stress component in the index would be inactivated ($\gamma=0$); otherwise, $\gamma$ would represent the degree of deviation from the threshold.

\begin{figure}
	\centering
	\includegraphics[width=4.5in]{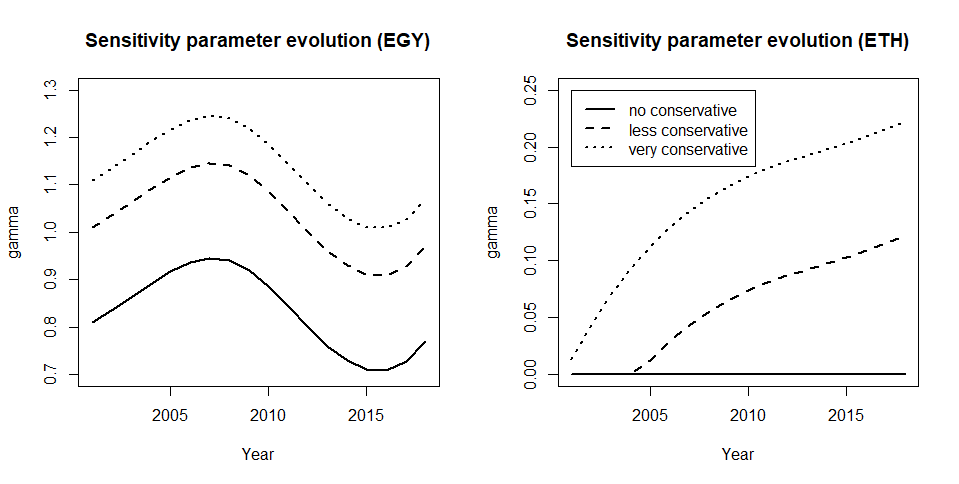}
	\caption{Evolution patterns for the sensitivity parameter $\gamma$ according to the perspectives stated in \eqref{gamma-rules} for Egypt and Ethiopia for the time period 2001-2018.}\label{fig-gamma-rules}
\end{figure}

For illustrative purposes, we adopt three distinct perspectives: (a) a very conservative (VC) stance, where $\gamma$ is adjusted when water stress exceeds the low-medium level ($>10\%$), (b) a less conservative (LC) approach, responding to breaches of the medium level ($>20\%$), and (c) a non-conservative (NC) viewpoint, which only considers significant deviations beyond the high water stress threshold ($>40\%$). These perspectives enable dynamic determination of the sensitivity parameter, reflecting the historical evolution of water stress and can be represented by the following rules:
\begin{equation}\label{gamma-rules}
	\left\{ \begin{array}{ll}
		\mbox{\emph{Very conservative perspective (VC):}} & \gamma_{c,t} := \max (0, W_{c,t-1} - 0.10)\\
		\mbox{\emph{Less conservative perspective (LC):}} & \gamma_{c,t} := \max (0, W_{c,t-1} - 0.20)\\
		\mbox{\emph{No conservative perspective (NC):}} & \gamma_{c,t} := \max (0, W_{c,t-1} - 0.40)
		\end{array} \right.
\end{equation}		
The calculated values for $\gamma$ under these three perspectives (VC, LC, NC) are depicted in Figure \ref{fig-gamma-rules} for Egypt and Ethiopia over the period 2001-2018. For Egypt, which is classified as an extremely high water-stressed country, all three rules result in strictly positive values for $\gamma$. These values evolve in a similar manner for each perspective, differing only in their horizontal positioning, which varies according to the specific rule applied. In the case of Ethiopia, the NC perspective assigns no positive value to $\gamma$, as the water stress index remains below the $40\%$ activation threshold throughout this period. Conversely, the LC perspective begins to assign positive values to $\gamma$ after 2004, exhibiting an upward trend. Under the VC perspective, $\gamma$ is allocated positive values for the entire period, also following an increasing trend.

\begin{figure}[ht!]
	\centering
	\includegraphics[width=4.5in]{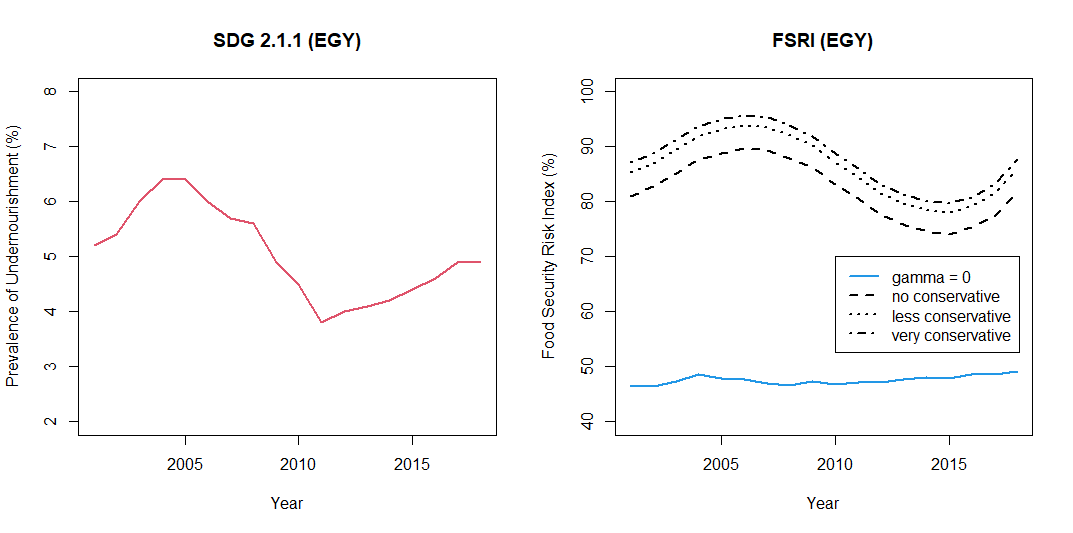}\\
	\includegraphics[width=4.5in]{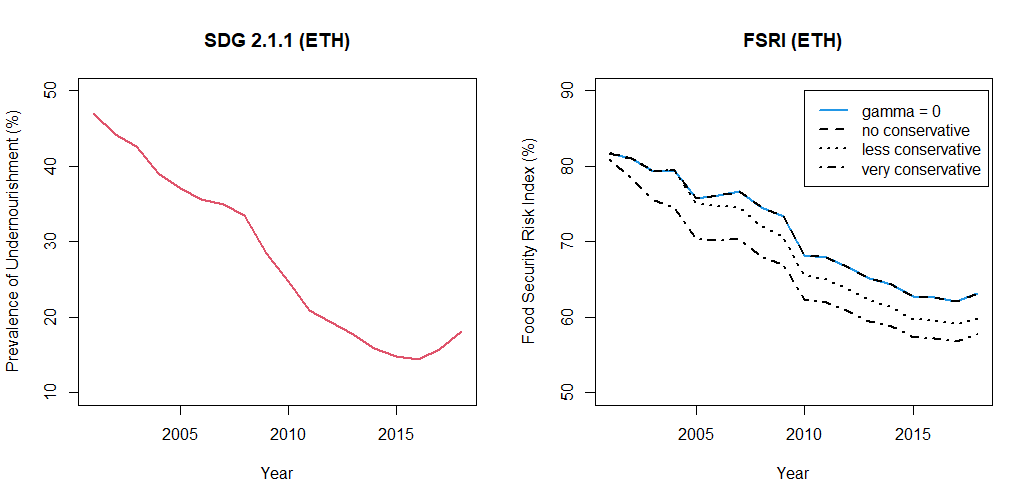}
	\caption{Illustration of the SDG 2.1.1 (PoU) index and FSRI for Egypt (upper panel) and Ethiopia (lower panel) for the period 2001 - 2018 under the perspectives stated in \eqref{gamma-rules}.}\label{fig-frsi-historical}
\end{figure}

While more intricate methods could be employed in aggregating the two risks, our straightforward, threshold-based approach seems to effectively approximate established indices like the PoU (SDG 2.1.1), which necessitate more detailed data. This is demonstrated in Figure \ref{fig-frsi-historical}, where we compare the PoU indicator with our proposed Food Security Risk Index (FSRI) using different values of $\gamma$ for both countries under study. For the sake of comparison, we also include scenarios where water stress is not factored into the food security risk calculation ($\gamma=0$). Notably, for Egypt, the trend of the PoU indicator is more accurately mirrored by the FSRI when positive values are assigned to $\gamma$, as opposed to the scenario excluding the water risk component ($\gamma=0$). In the case of Ethiopia, the FSRI closely approximates the trend of the PoU without significant discrepancies, regardless of the $\gamma$ value chosen. However, this alignment might shift if Ethiopia transitions into a highly water-stressed stage, at which point the water risk component of the index would likely become more influential in accurately depicting the food security risk.

Following the SSP-RCP scenarios framework introduced in Section \ref{POP-MODELS} combined with the modelling approaches discussed in Sections \ref{NEED-FOR-FOOD} and \ref{tfs-mod}, we are able to provide/generate probabilistic scenarios for the food security risk index $I^{FS}$ as follows: 
\begin{enumerate}
	
    \item {\bf Scenarios for Minimum Caloric Requirements}\\
    We use the procedure in Section \ref{NEED-FOR-FOOD} to provide future estimates and SSP compatible scenarios for the minimum calorie requirements at national level $C_{c,t}^{R}$ based on the detailed population evolution scenarios.
	
    \item {\bf Scenarios for Food System Capacity}\\
    {\bf (a)} Model layers \ref{upper-lm}-\ref{lower-lm} are calibrated using available historical data from a sufficiently large period for the country $c$ to estimate the profile of country $c$ (i.e. obtaining the relevant country-specific parameters). 
	
    {\bf (b)} The fitted model is then applied for predicting  the future food system capacity $Q^{FSC}_{c,t}$ using the probabilistic scenarios (and the relevant trajectories) for the population (see Section \ref{POP-SCENARIOS}) along with projections for the future GDP per capita as obtained from the global macroeconomic model MaGE \cite{foure2013, foure2021}. 
	
    \item {\bf Scenarios for the Food Security Risk Index}\\
    Using the trajectories for $Q^{FSC}_{c,t}$, $C_{c,t}^{R}$ and $W_{c,t}$ obtained in the previous steps we construct trajectories compatible with the various SSP-RCP scenarios for the index $I_{c,t}^{FS}$ and use the trajectories to provide statistical information for the index in the various scenarios. 
\end{enumerate}

Recalling our modelling framework (please see Figure \ref{fig-fsri-model}),  we realize that the generated values or trajectories for the food security risk indicator depend on the set of key factors
$$ Z_{c,t} := ( P^f_{a,c,t}, P^{m}_{a,c,t,} ,T_{c,t}, Pr_{c,t} )$$ . 
These  are the main stochastic factors which introduce uncertainty to the modelled quantities that lead to the estimation of $I^{FS}_{c,t}$. This relation is represented through a risk mapping $Z \mapsto I^{FS}(Z)$, connecting the random risk factors collected in $Z$ with the risk output $I^{SF}$, through the models presented in Sections \ref{NEED-FOR-FOOD} and \ref{tfs-mod} for a country $c$, i.e. expressing $I^{FS}$ in terms of $I^{FS} := \Phi_c(Z)$. Let us denote by $Q$ the probability measure under which the sample of trajectories for $Z$ is generated (i.e. coinciding with one of the SSP-RCP scenarios in Table \ref{tab-1.3}). From now on we will identify the output of the various scenarios, with respect to the risk factors $Z$, with probability measures $Q$, or equivalently with probability distributions for the risk factors, which in turn will induce  via the risk mapping $\Phi_c$ probability measures or probability distributions for the food security index $I^{FS}$. Employing the standard framework of risk management, we may use the above setting  in defining a risk measure associated with food security, as quantified by the index $I^{FS}$. In the case where a probabilistic model $Q$ for the description of the risk factors $Z$ is universally accepted, then, the best estimate for the risk would simply be the expectation of the risk mapping under the probability model $Q$, i.e. ${\mathbb E}_{Q}[I^{FS}]={\mathbb E}_{Q}[\Phi_{c}(Z)]$. In the present context, this would correspond to choosing one particular SSP-RCP scenario, the most plausible one, which would be identified with a probability measure $Q$ for the risk factors $Z$, and then using the risk mapping obtain an estimation of the risk measure in terms of ${\cal I}^{FS}={\mathbb E}[\Phi_c(Z)]$. Having more than one scenario, would correspond to obtaining a set of possible probability measures for the evolution of the risk factors $Z$, 
\begin{equation}\label{SSP-RCP-scenarios}
\begin{aligned}
\mathcal{Q} &=\{Q_1,Q_2, Q_3,Q_4,Q_5, Q_6\} \\
&=:\{ Q_{SSP_1-1.9}, Q_{SSP_1-2.6}, , Q_{SSP_2-4.5}, Q_{SSP_3-7.0}, Q_{SSP_4-6.0}, Q_{SSP_5-8.5}  \},
\end{aligned}
\end{equation}
and upon selecting any of these, the food risk security index is calculated as
\begin{equation}\label{fsr-within}
	\mathcal{I}^{FS}_{c,t} = \mathbb{E}_Q\left[ I^{FS}_{c,t} \right] = \mathbb{E}_Q\left[ \Phi_c(Z_{c,t}) \right].
\end{equation}

The methodology described above facilitates the estimation of food security risk within each of the considered socioeconomic-climate probabilistic scenarios, where `within'  implies basing our estimations on a singular scenario from the set $\mathcal{Q}$. However, identifying with certainty the exact scenario being experienced, particularly over long time horizons, is often not feasible due to the dynamic nature of the phenomena involved. In such instances, it becomes pragmatic to integrate various prospective scenarios to assess food security risk `across' scenarios. This approach reflects reality more closely, as the existence of a single, universally accepted probability model (scenario) $Q$ is rare. Instead, the set of probability models under different potential scenarios, $\mathcal{Q}$, offers diverse insights into the evolution of risk factors $Z$. This scenario leads us into the domain of Knightian uncertainty \cite{knight1921risk}, where there is no singular, universally acceptable probability measure for describing the evolution of risk factors affecting the phenomenon under study. In such contexts, deriving the best estimate for the risk extends beyond merely computing the expectation $\mathbb{E}_{Q}[\Phi_c(Z)]$ under a unique probability measure. This necessitates an estimator for the risk `across' scenarios – a measure of food security risk under the uncertainty of which scenario (equivalently, model) will actually materialize. One well-accepted approach in the risk management community within the context of \textit{convex risk measures} and their robust representation involves using the variational representation (see e.g. \cite{detlefsen2005, follmer2002convex, frittelli2002}). In this framework, we calculate the risk through its variational representation.
\begin{eqnarray}\label{RISK-VAR}
	\mathcal{I}^{FS}_{c,t} := \rho(I^{FS}_{c,t}) = \sup_{Q \in \mathcal{P}} \left\{ {\mathbb E}_{Q}[\Phi_c(Z_{c,t})] - a(Q) \right\},
\end{eqnarray} 
where the risk estimation $\mathcal{I}^{FS}$ is obtained in terms of the risk measure $\rho(I^{FS})$ referring to the "position" $I^{FS}$ and $a : \mathcal{P} \to {\mathbb R}_{+}$\footnote{\scriptsize $\mathcal{P}$ denotes the space of probability models that can describe the evolution of the random variable $Z$} is a (convex) penalty function in the space of probability models, which penalizes certain scenarios as extreme or improbable. The risk measure defined in expression \eqref{RISK-VAR} proposes as an estimation of food security risk, for which one cannot trust only a single scenario $Q$, the worst case expected risk over all probability models, properly weighted by the penalty function $\alpha(\cdot)$ which penalizes certain probability models as too extreme. The variational nature of formula \eqref{RISK-VAR} imparts a robustness to the proposed risk measure, as it no longer relies on a singular model for the occuring risk but rather on a weighted estimate encompassing the entire set $\mathcal{P}$ of plausible models. This set is defined by the selection of the penalization term. Recent studies by \cite{papayiannis2018} and \cite{petrakou2022} have linked the choice of the penalty function to the heterogeneity of these plausible models. Such a choice offers several benefits, including the potential for an analytical approximation of the risk measure. Furthermore, it provides an intriguing interpretation of the probability model used for risk estimation as the result of an expert consensus process, establishing a quantitative link between measures of risk and uncertainty.

Having adopted the fundamental conceptual framework of treating each scenario as a different probabilistic model (probability measure) for the risk factors $Z$ (mainly population factors in this study) we may now answer the question of robust estimation of the quantity of interest $I^{FS}_{c,t}$, for fixed $t$, using the convex risk measure of the form stated in \eqref{RISK-VAR}. Following the suggestion in \cite{papayiannis2018} and \cite{petrakou2022}, in order to distinguish between the various scenarios we adopt a metric space setting, according to which we consider the various scenarios in their natural setting, i.e., the metric space of probability measures, with a suitable metric $d(\cdot,\cdot)$, chosen so as to differentiate between the various probability measures in ${\cal Q}$, associated with different scenarios.

The risk measure is chosen so as to display uncertainty aversion, a property that is guaranteed (see \cite{petrakou2022}, see also \cite{papayiannis2018}) by choosing the penalty function
\begin{eqnarray}\label{P1}
	a(Q)= \frac{\theta}{2} \sum_{i=1}^{6} w_{i} d^{2}(Q, Q_{i}),
\end{eqnarray}
where $Q_{i}$ are the probability measures (probabilistic scenarios) included in the set $\mathcal{Q}$ stated in \eqref{SSP-RCP-scenarios}, where $w_{i}$, $i=1,2, \ldots, 6$ are (credibility) weights associated to each scenario (these can be subjective and associated to expert opinion or objective i.e. derived from evidence from the data and possibly updated through a suitable learning scheme), $d( \cdot, \cdot)$ is a metric in the space of probability measures and $\theta >0$ is the uncertainty aversion parameter, modelling the propensity of the decision maker to deviate from the probability models in ${\cal Q}$. A suitable choice  for $d$ is the Wasserstein metric (see e.g. \cite{santambrogio2015optimal, villani2021topics}) which is directly related to the misspecification error of the random variable $I^{FS}$ if a different probability model for $Z$ is chosen in place of the true model. Moreover, the choice \eqref{RISK-VAR}, with \eqref{P1} combined with the Wasserstein metric, allows for efficient numerical calculation of the risk measure ${\cal I}^{FS}$ for a wide class of probability models. For the interesting case where all plausible models are included in $\mathcal{Q}$ (corresponding to the limit $\theta \to \infty$) we obtain the approximation of ${\cal I}^{FS}$ by
\begin{eqnarray}\label{fsr-between}
	{\cal I}^{FS}_{c,t} =  \mathbb{E}_{Q^{*}}[I^{FS}_{c,t}] = {\mathbb E}_{Q^{*}} [\Phi_c( Z)],
\end{eqnarray}
where $Q^{*}$ is the barycentric probability model over all scenarios in $\mathcal{Q}$ with respect to the weights $w$ (under the Wasserstein distance sense, see e.g. \cite{agueh2011barycenters}). The risk measure ${\cal I}^{FS}$ as stated above, provides a robust estimate for the food security risk, across scenarios, in the limit of deep uncertainty. This is the main approximation we will be using in this work, however further approximations are possible, if required, using the $\Delta$-approximation of the risk mapping $Z \to \Phi_c( Z)$ (for details please see \cite{papayiannis2018}).


The algorithmic approach in estimating the food security risk index across scenarios can be thus summarized as follows:

\begin{algorithm}[Food Security Risk Estimation Across SSP-RCP scenarios]\hfill 
\begin{enumerate}
	\item Fix a certain country $c$ and a time $t$.
	\item Define the risk factors $Z=(P_{a,f},P_{a,m}, T, Pr)$ and using the procedure described in Section \ref{POP-SCENARIOS}  obtain probabilistic scenarios for $Z$ and determine the corresponding probability models $Q_{i}$, $i=1, \ldots, 6$ through the generated samples.
	\item By estimating the model layers described in \eqref{upper-lm}-\eqref{lower-lm} and using the expression \eqref{TOTAL-CALORY-INTAKE} obtain the risk mapping $Z \mapsto I^{FS}_c = \Phi_c(Z)$ for $I^{FS}_c$ as defined in \eqref{fsri}.
	\item Obtain the barycentric scenario $Q^{*}$ provided a choice of weights $w$ allocating the degree of realism of the policy maker to each one of the scenarios under consideration.
	\item Using Monte-Carlo simulation estimate ${\cal I}_c^{FS}= {\mathbb E}_{Q^{*}} [\Phi_c(Z)]$.
\end{enumerate}
\end{algorithm}

\section{Food Security Risk in Egypt and Ethiopia}\label{sec-4}

\subsection{Data and model assumptions}

Adhering to the modeling approach presented in Section \ref{model}, for (a) the minimum caloric requirements (Section \ref{NEED-FOR-FOOD}) and (b) the food system capacity (Section \ref{tfs-mod}), in this section we estimate the food security risk in Egypt and Ethiopia, as per the risk indicator introduced in Section \ref{food-security-indices}. For estimating the minimum calorie intake component, we primarily utilize probabilistic projections of population structure for both countries, as this quantity is predominantly dependent on population (refer to Section \ref{NEED-FOR-FOOD} for details). On the other hand, the food system capacity at the country level is delineated through the two-layer model described in \ref{upper-lm}-\ref{lower-lm}, necessitating relevant historical data for model calibration. The calibration task employed data from the 1990-2018 period, during which comprehensive data for all relevant quantities were available.  The data for model fitting for Egypt and Ethiopia were principally sourced from the FAO's open database\footnote{\scriptsize \url{https://www.fao.org/faostat/en/\#data}}, which provides historical information on food system capacity, food quantity imports and exports, and land area used for agricultural purposes at the national level. Climate data, specifically average temperature and precipitation, were obtained from the World Bank's Climate database\footnote{\scriptsize \url{https://climateknowledgeportal.worldbank.org/}}, while water-stress records and estimations were sourced from the World Bank Database\footnote{\scriptsize \url{https://data.worldbank.org/}}. The variables utilized in the model calibration and the sources of the corresponding data are detailed in Table \ref{tab-train-data} in the Appendix \ref{App-C}.

For the model fitting procedure, we employed Cobb-Douglas type parametric models to estimate all necessary parameters. A comprehensive description of the full model is available in the Appendix \ref{App-C}. The fitting and estimation process was conducted separately for each country (not as panel data), incorporating Ridge-type penalization into the typical OLS estimation method. This approach was chosen to address potential instabilities arising from the collinearity effects of the predictors. It is also important to note that models were fitted individually to data, rather than as a system of equations, due to the consideration of different sets of effects for each quantity. The resulting model parameters for Egypt and Ethiopia are presented in Tables \ref{fsc-mod-EGY} and \ref{fsc-mod-ETH} in the Appendix \ref{App-C}, along with the corresponding goodness-of-fit results.  In terms of the food system capacity profiles, there are no significant qualitative differences observed between the two countries. The technological effects and the year trend are comparable for domestic food production. A notable qualitative distinction is seen in the effect of food exports for Egypt, where the relevant coefficient is negative, possibly reflecting differences in food production capabilities and patterns between the two countries. Models for exported and imported quantities appear quite similar for both Egypt and Ethiopia. Notably, Egypt's food production mechanism shows a negative correlation with potential temperature increases. In Ethiopia, a negative relationship is observed between the labor force in food production activities and GDP, hinting at potential social trends with increasing income. However, these effects should be interpreted as aggregate influences, and it would be an intriguing area for future research to examine if these effects persist in more granular models, such as those differentiating between food products or regions. Such an endeavor, however, hinges on the availability of more detailed data, which is currently not accessible.

\subsection{Food security risk within SSP-RCPs}%

Following the modelling of minimum food requirements and food system capacity, estimates for both quantities concerning the time period 2019-2050 for Egypt and Ethiopia under each SSP-RCP scenario are produced (for the list of scenarios presented in Table \ref{tab-1.3}). Population projections under each SSP scenario are directly available through the probabilistic model presented in Section \ref{POP-SCENARIOS}. Concerning the other socioeconomic indicators, there exist several macroeconomic models that, based on population projections, provide projections for various socioeconomic indicators of interest under the SSP scenarios. In this paper, the MaGE model (see Section \ref{POP-SCENARIOS} and \cite{foure2013, foure2021} for more details) is employed for this task. The MaGE model based on the UN and IIASA\footnote{\scriptsize \url{https://iiasa.ac.at/}} databases provides projections up to year 2100 for the socioeconomic activities of all countries of the world (being a global model) under each one of the SSP scenarios. In this paper, the MaGE model is extended in order to provide projections of the socioeconomic variables that are compatible with the probabilistic population projections. This is achieved by substituting the standard population inputs in MaGE with the conditional means for population under each SSP, obtained in terms of the simulated trajectories (accordingly assigned to the appropriate SSP scenario using the methodology introduced in Section \ref{POP-SCENARIOS}), to provide estimates for GDP and labour, compatible with the philosophy of SSP probabilistic modelling. Note that this task could be performed trajectory-wise from the population scenario database i.e., generating individual trajectories using MaGE for the socioeconomic quantities in question, one for each individual trajectory for the population evolution, within scenarios. This would lead to a sample of trajectories per scenario for the socioeconomic quantities, which could then be used for the generation of probabilistic projection for the socioeconomic quantities, in a manner similar to that described in Sections \ref{POP-MODELS} and \ref{model}. However this would be quite expensive in computational time and possibly colliding with the general modelling philosophy of MaGE which is based on pointwise estimates, therefore it would essentially require a brand new macroeconomic model.

For the projection tasks of this work, these two worlds are combined using the future estimates for GDP evolution as provided by MaGE under each one of the SSP scenarios, incorporating the conditional means for the population related quantities as obtained by the full probabilistic population model, keeping in mind the possible limitations and drawbacks from this approach. Climate drivers and in particular temperature and precipitation (in annual basis), are provided for each one of the scenarios in Table \ref{tab-1.3} from the Climate Change Knowledge Portal online database. Projections about the land area that will be occupied for agricultural activities, are performed under the assumption that no changes to the current land use policies are made. However, constraints as to the maximum area occupied for agricultural purposes according to each country capabilities in providing agricultural land, are applied. In this direction, for Egypt and Ethiopia we assign a maximum land area value that can be allocated to agriculture until 2050 with respect to the current capabilities of the countries. Therefore, to determine each country's profile in agricultural land use, we estimated its trend for the last two decades, restricted by the upper bound that has been set by the physical limitations in land use. In this way, since land use depends on population, economy and climate factors (see model \eqref{lower-lm}), different estimates are obtained under each SSP-RCP scenario. Although more sophisticated models can be considered, taking into account thinner data about land use, under which the effect of land policies might be assessed by the proposed approach, we believe that such a task is beyond the scopes of this work. The key socio-economic and climate drivers which where used for the projections and the related models and sources that they were obtained from, are illustrated in Table \ref{tab-proj-data} in Appendix \ref{App-C}.

After modeling the minimum food requirements and food system capacity, we proceed to produce estimates for both quantities for the period 2019-2050 for Egypt and Ethiopia under each SSP-RCP scenario listed in Table \ref{tab-1.3}. Population projections under each SSP scenario, as presented in Section \ref{POP-SCENARIOS}, are readily available from the probabilistic model. For other socioeconomic indicators, several macroeconomic models, including the MaGE model (discussed in Section \ref{POP-SCENARIOS} and \cite{foure2013, foure2021}), provide projections based on these population forecasts under the SSP scenarios. The MaGE model, utilizing data from UN and IIASA\footnote{\scriptsize \url{https://iiasa.ac.at/}}, offers global projections up to 2100 for socioeconomic activities across all countries for each SSP scenario. 

Although it is theoretically possible to generate individual trajectories for socioeconomic quantities using MaGE for each population trajectory within scenarios, this approach would be computationally intensive and potentially conflict with MaGE's focus on pointwise estimates, necessitating a new macroeconomic model. In this study, we generate projections for socioeconomic variables that align with the probabilistic population projections by substituting the standard population inputs in MaGE with the conditional means for population under each SSP, derived from the simulated trajectories appropriately assigned to each SSP scenario as described in Section \ref{POP-SCENARIOS}. This approach provides estimates for GDP and labor that are consistent with the SSP probabilistic modeling philosophy. However, limitations and drawbacks of this approach are acknowledged. 

Climate drivers, specifically annual temperature and precipitation, for each scenario in Table \ref{tab-1.3}, are sourced from the Climate Change Knowledge Portal online database. Projections for agricultural land use assume no changes in current land policies, with constraints applied based on each country's capacity for agricultural land provision. For Egypt and Ethiopia, we set a maximum agricultural land area up to 2050, estimating its trend over the last two decades within physical land use limitations. This results in different agricultural land use estimates under each SSP-RCP scenario, based on population, economy, and climate factors (see model \eqref{lower-lm}). While more intricate models could be used to analyze detailed land use data, this is outside the scope of this study. The key socio-economic and climate drivers used for these projections, along with the models and sources from which the data were obtained, are detailed in Table \ref{tab-proj-data} in the Appendix \ref{App-C}.

\begin{figure}[ht!]
	\centering
	\includegraphics[width=5in]{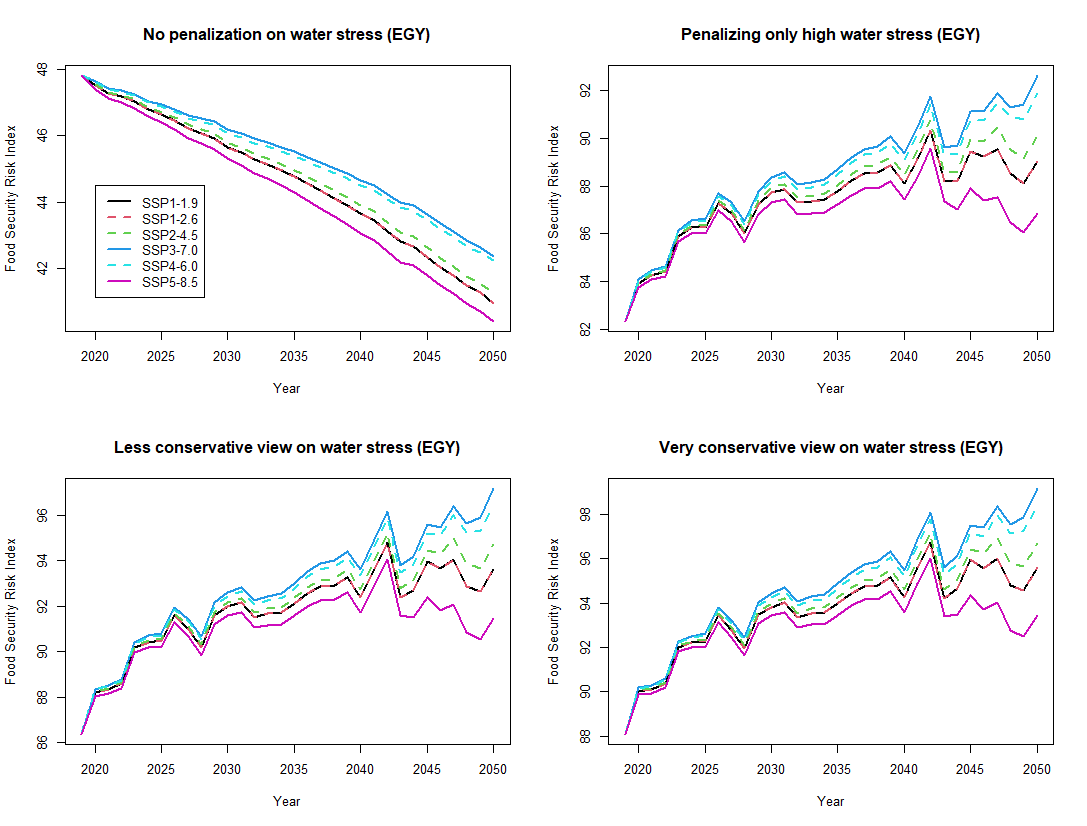}
	\caption{Graphical illustration of the food security risk index (median trajectories) for Egypt for the time period 2019-2050 under each SSP-RCP scenario and water-stress risk aggregation perspectives.}\label{fig-fsi-wsi-EGY}
\end{figure} 

Integrating the projections discussed previously, we obtain estimates for food security under each SSP-RCP scenario for Egypt and Ethiopia, using the food security index outlined in Section \ref{food-security-indices}. These estimates are derived under four different perspectives for setting the sensitivity parameter $\gamma$: (a) $\gamma=0$ (excluding water stress), (b) a VC perspective ($\gamma>0$ if $W_{c,t-1}>0.10$), (c) a LC perspective ($\gamma>0$ if $W_{c,t-1}>0.20$), and (d) a NC perspective ($\gamma>0$ if $W_{c,t-1}>0.40$). According to the findings in Figures \ref{fig-fsi-wsi-EGY}, the food security risk for Egypt is projected to increase in the coming decades, with a higher rate initially and a slower rate after 2040 under most SSP-RCP scenarios. A notable exception is a slight decrease in risk post-2040 under the SSP5-8.5 scenario. This trend is consistent across all methods of selecting $\gamma$, except when the water risk component is omitted ($\gamma=0$). Given Egypt's status as a highly water-stressed country—a situation unlikely to change soon—the exclusion of water risk ($\gamma=0$) yields an unrealistic portrayal. Water resource availability has long been a critical issue in North Africa, particularly in Egypt, and remains unresolved. Consequently, the depiction by the index for $\gamma=0$ does not align with reality, considering the strong dependency of the Egyptian agricultural sector on water resources (see, for example, \cite{christoforidou2023food, osman2016water} and references therein), and the significant impact of water scarcity hazards on food production, notably in cereal production.

\begin{figure}[ht!]
    \centering
    \includegraphics[width=5in]{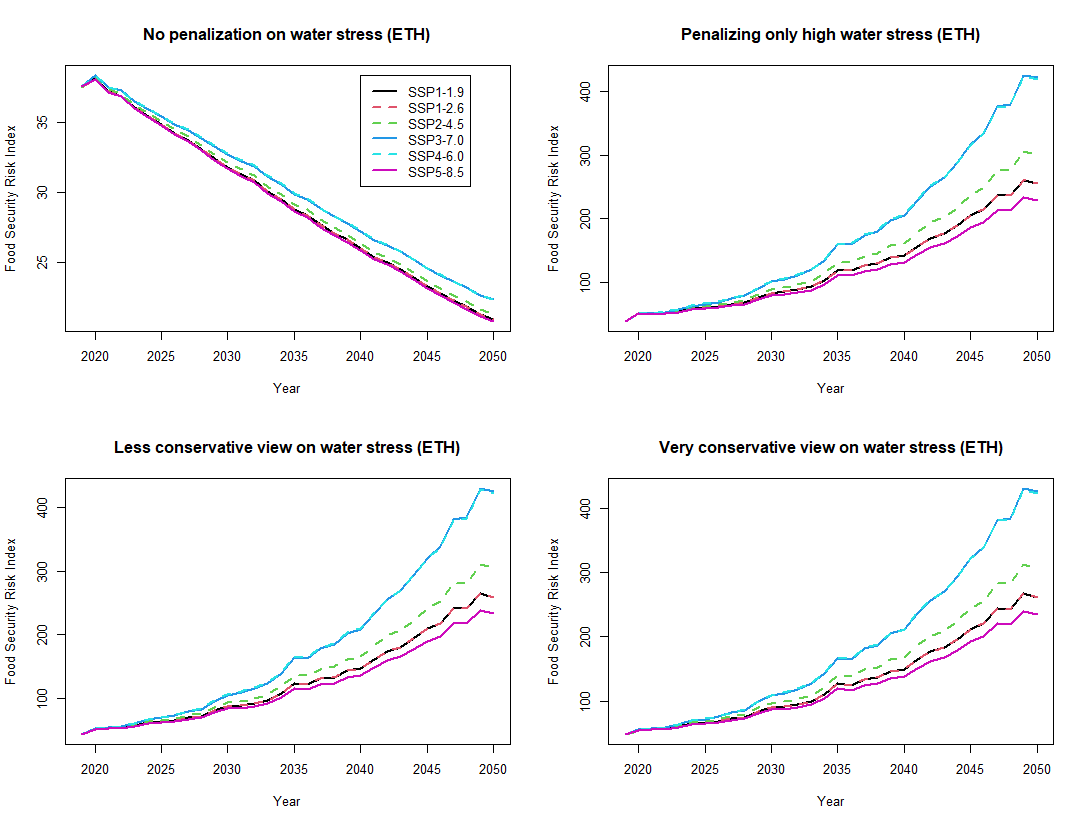}
	\caption{Graphical illustration of the food security risk index (median trajectories) for Ethiopia for the time period 2019-2050 under each SSP-RCP scenario and water-stress risk aggregation perspectives.}\label{fig-fsi-wsi-ETH}
\end{figure}

For Ethiopia, the estimated food security risk, when water stress is not aggregated into the calculation ($\gamma=0$), shows a similar pattern (see Figure \ref{fig-fsi-wsi-ETH}). Under all SSP-RCP scenarios, the risk appears to decrease markedly, almost exponentially. However, Ethiopia's situation differs significantly from Egypt's. Despite currently not being classified as a water-stressed country (with the water stress indicator below 40\%), the scenario for Ethiopia is expected to shift dramatically in the coming years, considering the increasing trend in historical water stress records. Our water stress model projects that by 2050, Ethiopia's median water stress will likely range between 300\% to 500\% under all considered scenarios (refer to Figure \ref{fig-wsi-EGY-ETH} in the Appendix \ref{App-C}). This implies that Ethiopia would need to secure an additional 3-5 times the current amount of water resources, a situation poised to significantly impact food security risk and, by extension, food production activities. Employing the water stress penalization perspectives from Section \ref{food-security-indices}, we gain a more accurate view of Ethiopia's future food security risk. Notably, even under the NC perspective, a considerable increase in food security risk is anticipated in the forthcoming years across all SSP-RCP scenarios due to the predicted water scarcity hazard. This trend holds true under all perspectives (NC, LC, and VC), indicating a substantial food security risk by 2050, particularly notable in the mid-range SSP2-4.5 scenario.

\subsection{Food security risk across SSP-RCPs}

While obtaining food security risk estimates within each SSP scenario is insightful, a key limitation of this approach is its lack of robustness to uncertainties regarding which scenario will actually materialize. To address this, and aid policymakers in making informed decisions that consider all possible scenarios, we implement the robust risk assessment approach discussed in Section \ref{food-security-indices}. Illustrating this with a toy example, we consider three hypothetical types of policymakers (or experts\footnote{\scriptsize For insights into incorporating expert opinions, see \cite{sutherland2015policy}}): (a) one characterized by complete ignorance, hence assigning equal weight to all possible outcomes; (b) an optimist, who assigns higher probabilities to scenarios more favorable for the environment; and (c) a pessimist, leaning towards scenarios less favorable for the environment. These types are detailed in Table \ref{tab-7}.

\begin{table}[ht!]\small
	\centering
	\begin{tabular}{l|cccccc}
		\hline\hline
		\bf Expert's Opinion & \bf SSP1-1.9 & \bf SSP1-2.6 & \bf SSP2-4.5 & \bf SSP3-7.0 & \bf SSP4-6.0 & \bf SSP5-8.5\\
		\hline
		A. Ignorance    & 1/6    & 1/6   & 1/6  & 1/6   & 1/6   & 1/6\\
		B. Optimistic 	& 1/2    & 1/5   & 3/20 & 1/25 & 1/10   & 1/100\\
		C. Pessimistic 	& 1/100  & 1/25  & 1/10 & 1/5   & 3/20   & 1/2\\
		\hline\hline
	\end{tabular}
	\caption{Realization probabilities of each SSP-RCP scenario according to different perspectives}\label{tab-7}
\end{table}


\begin{figure}[ht!]
\centering
\includegraphics[width=6.2in]{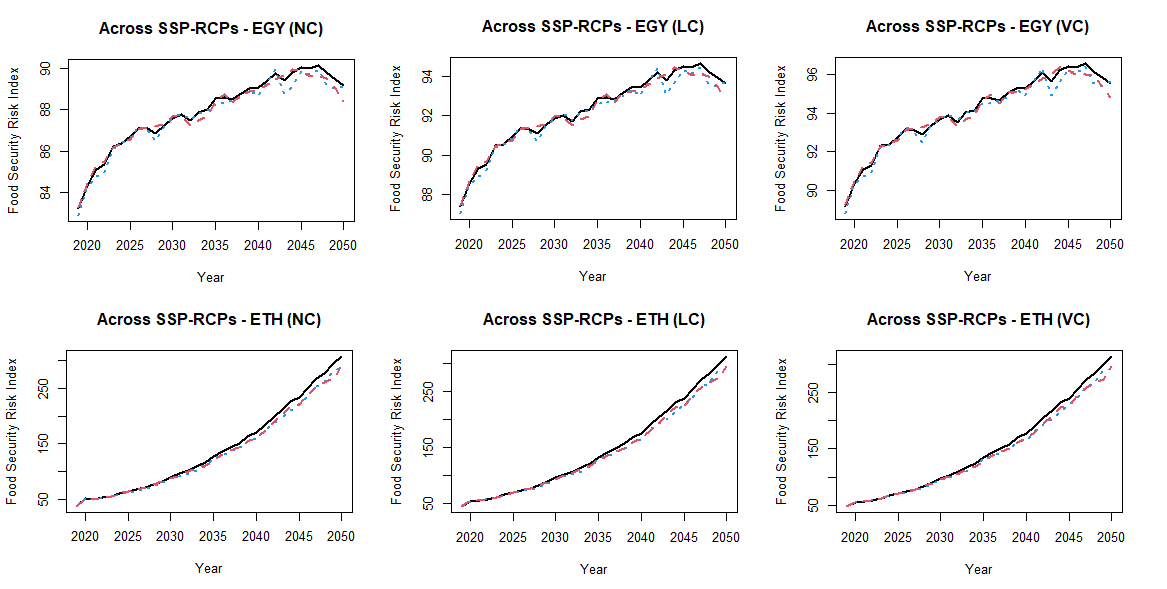}
\caption{Graphical illustration of the across scenario estimations of the food security index (left column) and water stress index (right column) for Egypt (upper panel) and Ethiopia (lower panel) for the time period 2019-2050.}\label{fsri-robust}
\end{figure}

Figure \ref{fsri-robust} presents the estimates of the food security index across SSP-RCP scenarios, applying the NC, LC, and VC rules for selecting $\gamma$. For Egypt, the estimations under all water stress perspectives align closely, showing a consistent trend across different expert opinions. This trend indicates an increasing concave-type risk in food security up to the year 2040, followed by a slight decline. This pattern is largely attributed to Egypt's high vulnerability to water scarcity. In the case of Ethiopia, the food security risk estimates also exhibit homogeneity across all perspectives and expert types, revealing an almost exponential increase in food security risk over time. This robust scenario aggregation approach suggests that both countries are facing significant food security challenges, with Ethiopia's situation predicted to deteriorate more rapidly unless preemptive actions are implemented. Importantly, these estimates across scenarios for both countries demonstrate their robustness, offering consistent predictions that appear relatively unaffected by variations in weights assigned to different SSP-RCP scenarios (reflecting various types of expert opinions).

\section{Conclusions}\label{sec-5}

In this work, we proposed a general methodology for creating probabilistic socioeconomic scenarios aligned with the SSP-RCP framework and for assessing food security risk. This methodology stands out for its emphasis on probabilistic scenarios, which effectively represent the inherent uncertainty and are more suitable for projecting key socioeconomic quantities into the future. As an application, we focused on food security risk, providing a plausible index for assessment and a quantitative evaluation framework per scenario, along with future projections. Acknowledging the uncertainty in which scenario materializes, we also introduced a robust framework for creating projections across scenarios using the concept of convex risk measures in the presence of model uncertainty.

The application of this approach to Egypt and Ethiopia in the upper Nile region reveals diverging food security risks. While Egypt is not expected to face significant food security risk over the next 30 years under any SSP-RCP trajectory, Ethiopia is predicted to experience an exponential increase in water stress, potentially requiring 3-5 times its current water resources by 2050. This alarming projection should urge Ethiopian policymakers to design sustainable water use policies to prevent severe food security issues. Our analysis remains robust across different hypotheses about scenario probabilities, emphasizing the urgent need for Ethiopian policy-makers to explore pathways for food system transformation, as suggested by \cite{who2021}.

However, the various models employed in this paper are subject to certain assumptions and limitations. The Raftery's population model and the MaGE, which form the basis of our probabilistic scenario modeling, do not explicitly consider population control policies or climate effects. These are introduced exogenously through RCP scenarios. Future extensions could explore the explicit integration of these effects. Additionally, the data used for calibrating the food capacity model only cover the period up to 2018, excluding recent phenomena like pandemics or political events. Techniques from extreme value theory could be employed to assess the impacts of these shocks.

Finally, while a more detailed food security model incorporating local food production infrastructure, dietary patterns, etc., could be developed, the lack of reliable data for many countries necessitates a more aggregate approach. Thus, our proposed methodology and the resulting food security risk index rely on publicly available, accessible, and reliable data. This approach offers a balanced perspective, providing valuable insights for policymakers while acknowledging the constraints and potential areas for future research.

\section*{Acknowledgements}

The authors would like to thank the Editor, the Associate Editor and the two anonymous referees for their insightful, constructive comments and suggestions that helped in improving the quality of the manuscript.

\section*{Funding}
This project has received funding from the European Union's Horizon 2020 partnership for research and innovation in the Mediterranean area programme (PRIMA) under grant agreement No 1942 (AWESOME).

\newpage

\appendix

\section{Extra Material from Section \ref{POP-MODELS}}\label{App-A}

\subsection*{Population evolution rates projections for Egypt and Ethiopia under the probabilistic setting}

\begin{figure}[ht!]
\centering
\includegraphics[width=5.5in]{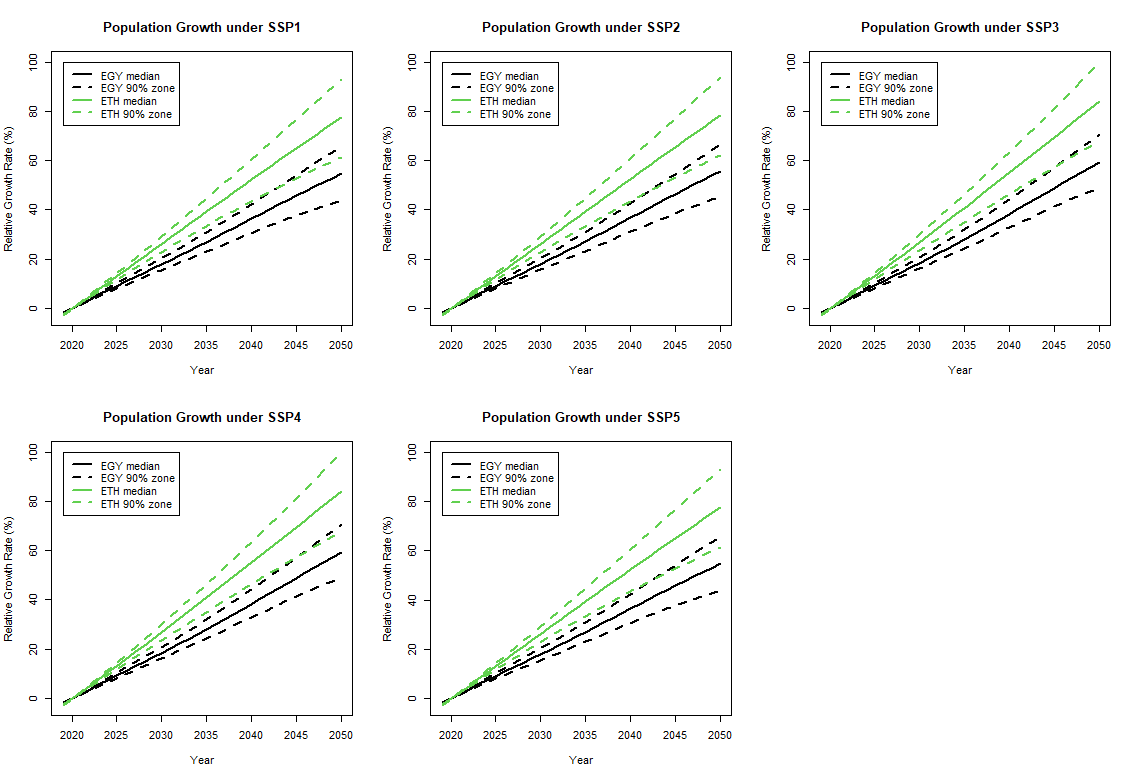}
\caption{Relative population growth median projections and 90\% zones for Egypt and Ethiopia under each SSP scenario using 2020 as the reference year.}\label{EGY-ETH-unc}
\end{figure}

\subsection*{RCP Scenarios Specifications}

\begin{table}[ht!]
\centering
\begin{tabular}{llll}
\hline\hline
\bf Scenario    &  \bf Emissions Level  & \bf Temperature Change & \bf Mitigation Measures \\
\hline
RCP1.9   & Best-case         & between $1-1.5^o$C      & Extremely stringent \\
RCP2.6   & Low                 & between $1.5-2^o$C      &  Very stringent         \\ 
RCP4.5   & Medium - Low  & between $2.5-3^o$C      &  Less stringent         \\
RCP6.0   & Medium - High & between $3-3.5^o$C      &  Very loose  \\
RCP7.0   & High                & up to $4^o$C until 2100 &  Extremely loose  \\ 
RCP8.5   & Worst-case      & up to $5^o$C until 2100  &  No mitigation \\                         
\hline\hline                           
\end{tabular}
\caption{Representative Concentration Pathways (RCP) definitions}\label{tab-1.2}
\end{table}

\newpage
\section{Uncertainty Quantification Issues}\label{App-B}

\begin{figure}[ht!]
\centering
\begin{subfigure}{0.48\textwidth}
\includegraphics[width=\textwidth]{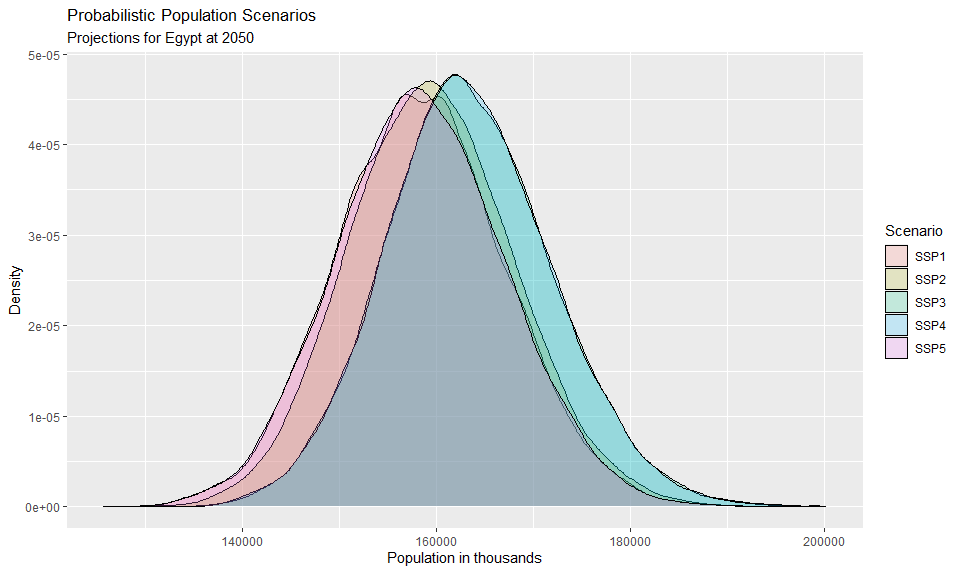}
\caption{Total population}
\end{subfigure}
\hfill
\begin{subfigure}{0.48\textwidth}
\includegraphics[width=\textwidth]{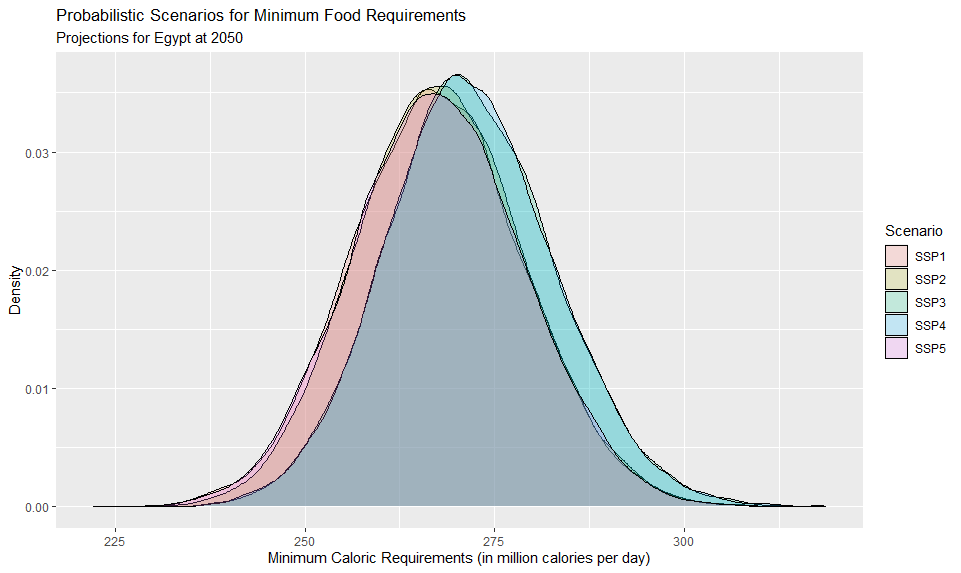}
\caption{Minimum caloric requirements}
\end{subfigure}
\hfill
\begin{subfigure}{0.48\textwidth}
\includegraphics[width=\textwidth]{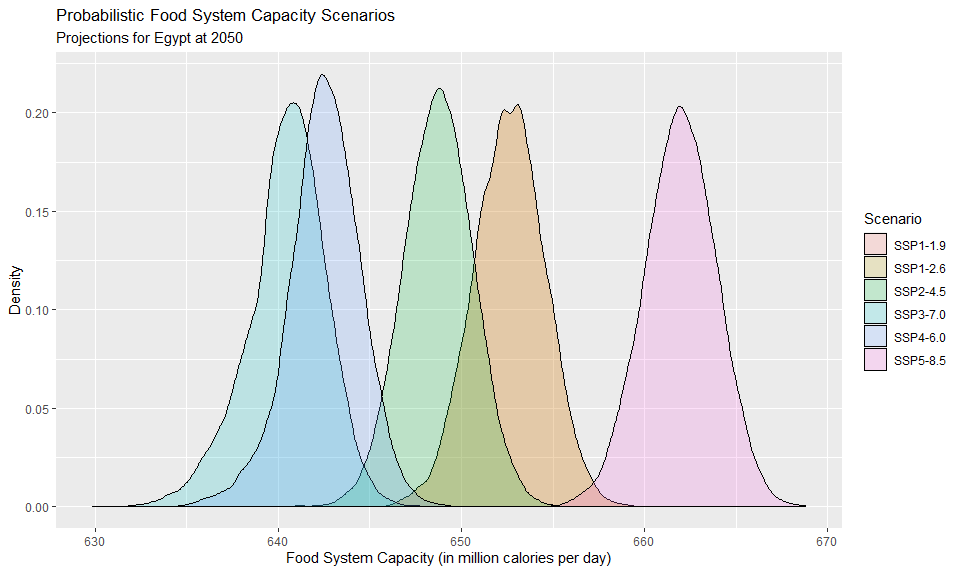}
\caption{Food system capacity}
\end{subfigure}
\hfill
\begin{subfigure}{0.48\textwidth}
\includegraphics[width=\textwidth]{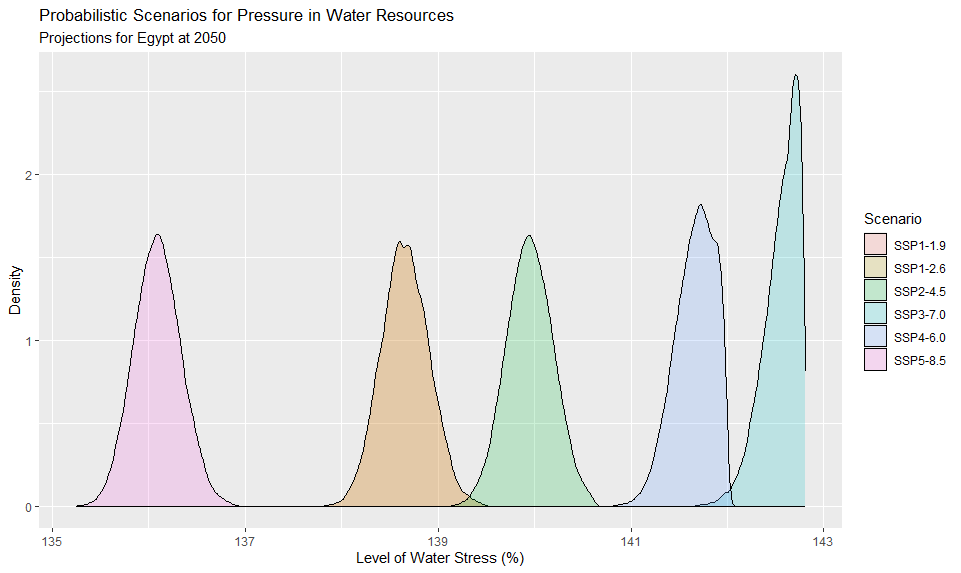}
\caption{Level of water stress}
\end{subfigure}
\hfill
\begin{subfigure}{0.48\textwidth}
\includegraphics[width=\textwidth]{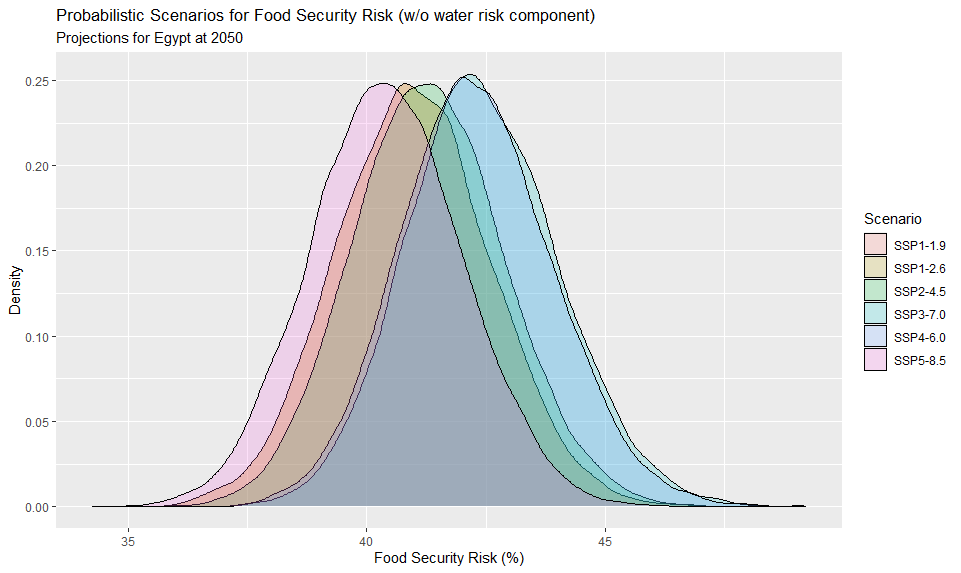}
\caption{Food security risk (no water risk)}
\end{subfigure}
\hfill
\begin{subfigure}{0.48\textwidth}
\includegraphics[width=\textwidth]{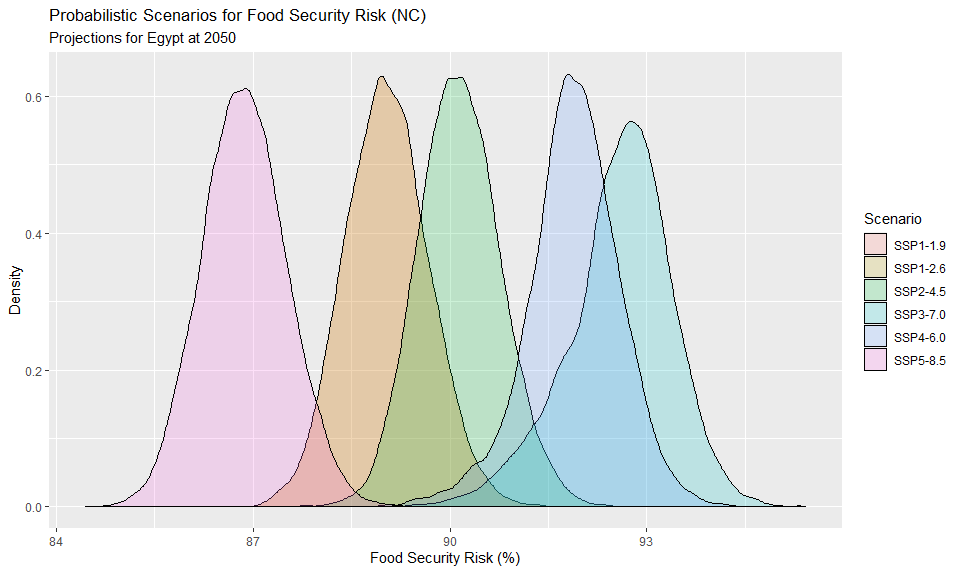}
\caption{Food security risk (NC)}
\end{subfigure}
\hfill
\begin{subfigure}{0.48\textwidth}
\includegraphics[width=\textwidth]{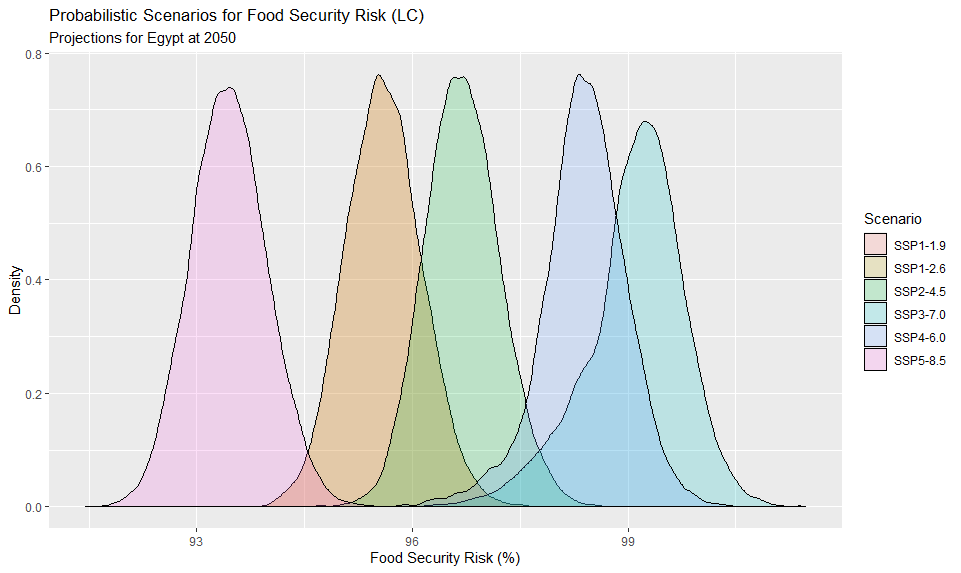}
\caption{Food security risk (LC)}
\end{subfigure}
\hfill
\begin{subfigure}{0.48\textwidth}
\includegraphics[width=\textwidth]{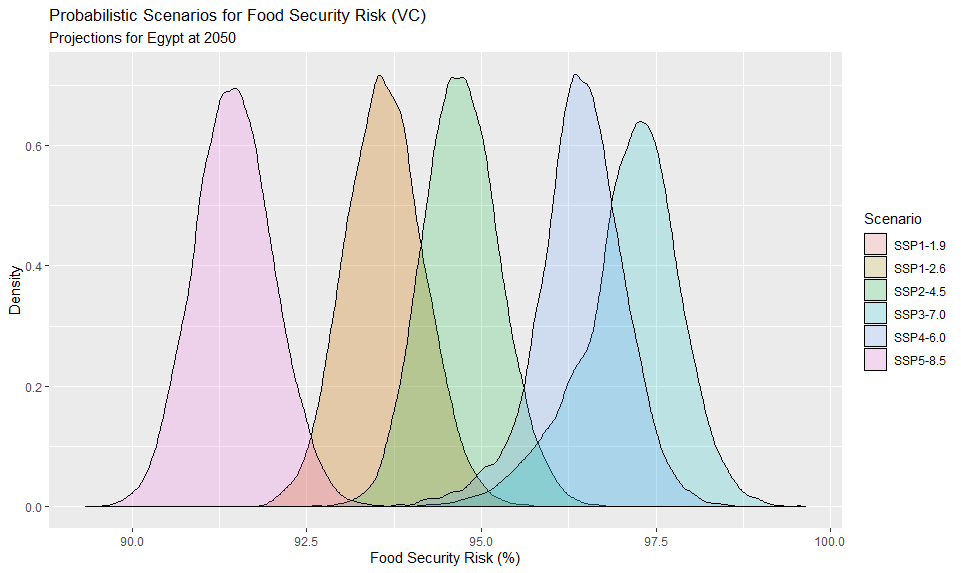}
\caption{Food security risk (VC)}
\end{subfigure}

\caption{Probabilistic projections for Egypt at the year 2050 for food security risk and relevant key quantities}\label{prob-proj-EGY}
\end{figure}

\begin{figure}[ht!]
\centering
\begin{subfigure}{0.48\textwidth}
\includegraphics[width=\textwidth]{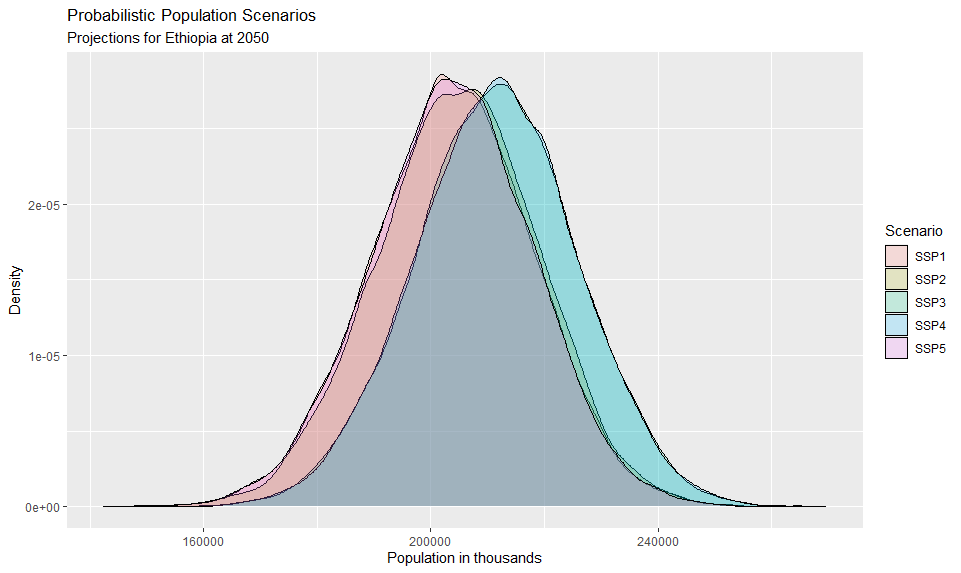}
\caption{Total population}
\end{subfigure}
\hfill
\begin{subfigure}{0.48\textwidth}
\includegraphics[width=\textwidth]{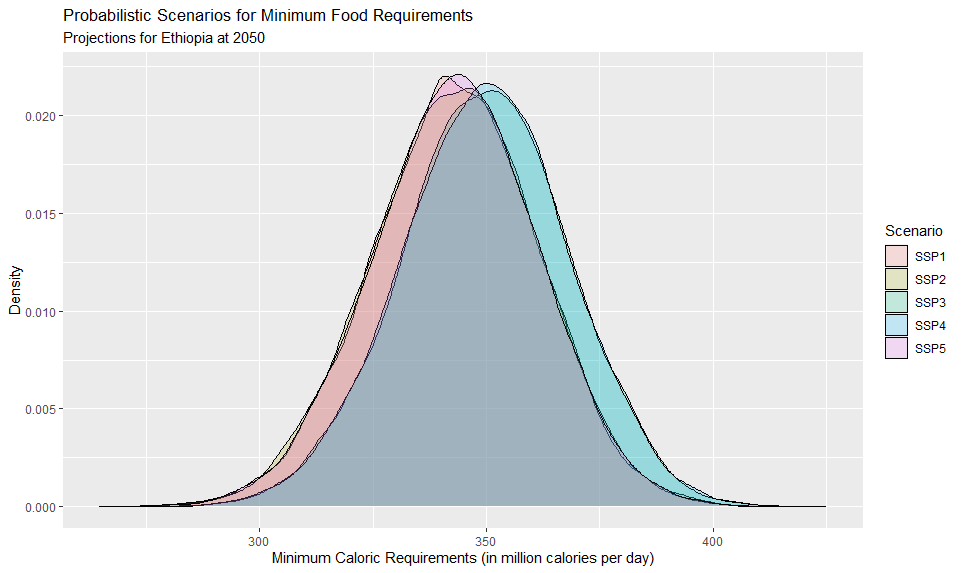}
\caption{Minimum caloric requirements}
\end{subfigure}
\hfill
\begin{subfigure}{0.48\textwidth}
\includegraphics[width=\textwidth]{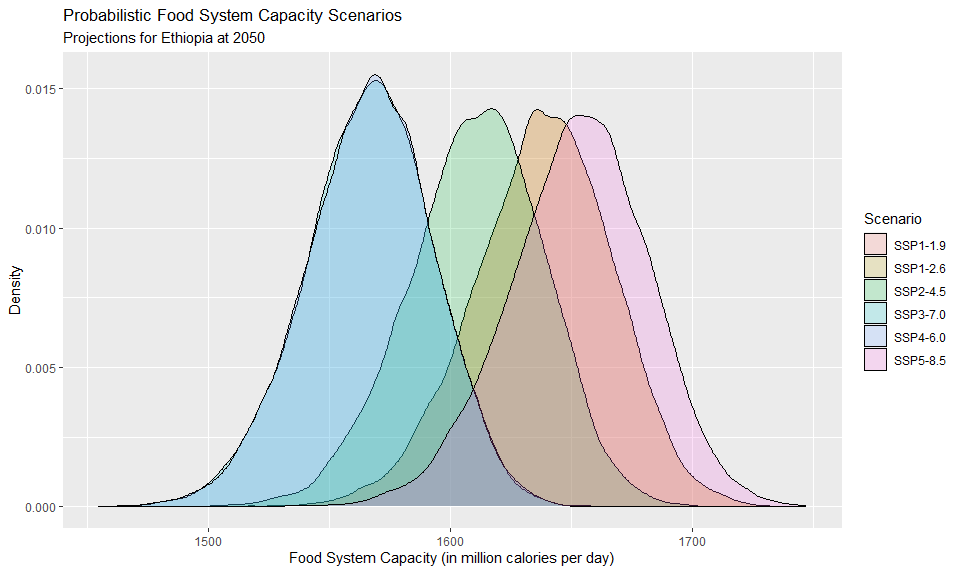}
\caption{Food system capacity}
\end{subfigure}
\hfill
\begin{subfigure}{0.48\textwidth}
\includegraphics[width=\textwidth]{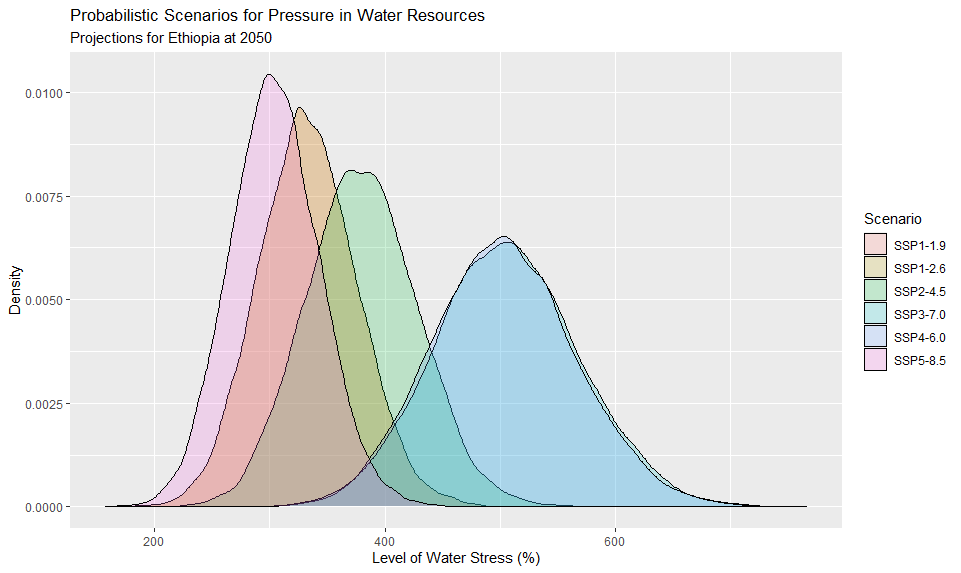}
\caption{Level of water stress}
\end{subfigure}
\hfill
\begin{subfigure}{0.48\textwidth}
\includegraphics[width=\textwidth]{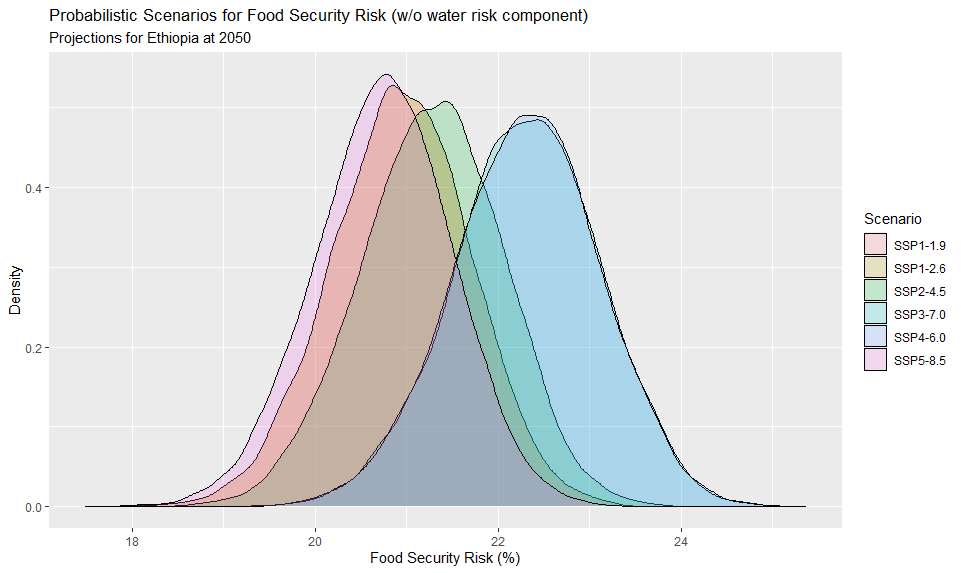}
\caption{Food security risk (no water risk)}
\end{subfigure}
\hfill
\begin{subfigure}{0.48\textwidth}
\includegraphics[width=\textwidth]{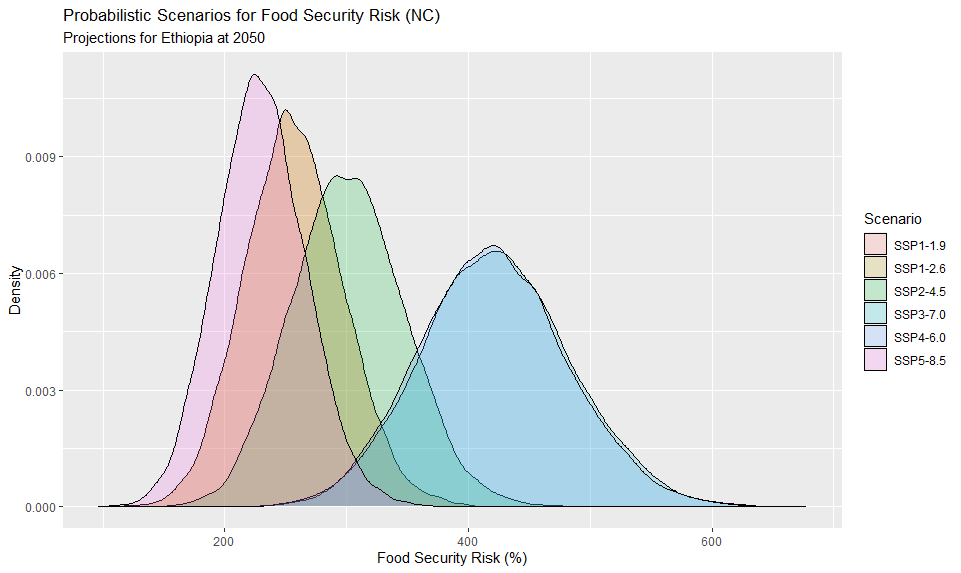}
\caption{Food security risk (NC)}
\end{subfigure}
\hfill
\begin{subfigure}{0.48\textwidth}
\includegraphics[width=\textwidth]{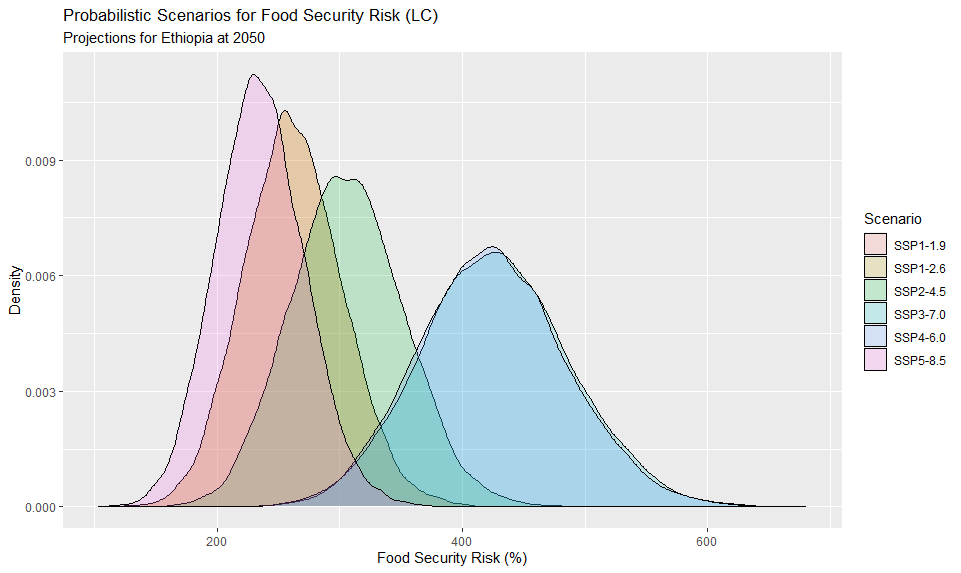}
\caption{Food security risk (LC)}
\end{subfigure}
\hfill
\begin{subfigure}{0.48\textwidth}
\includegraphics[width=\textwidth]{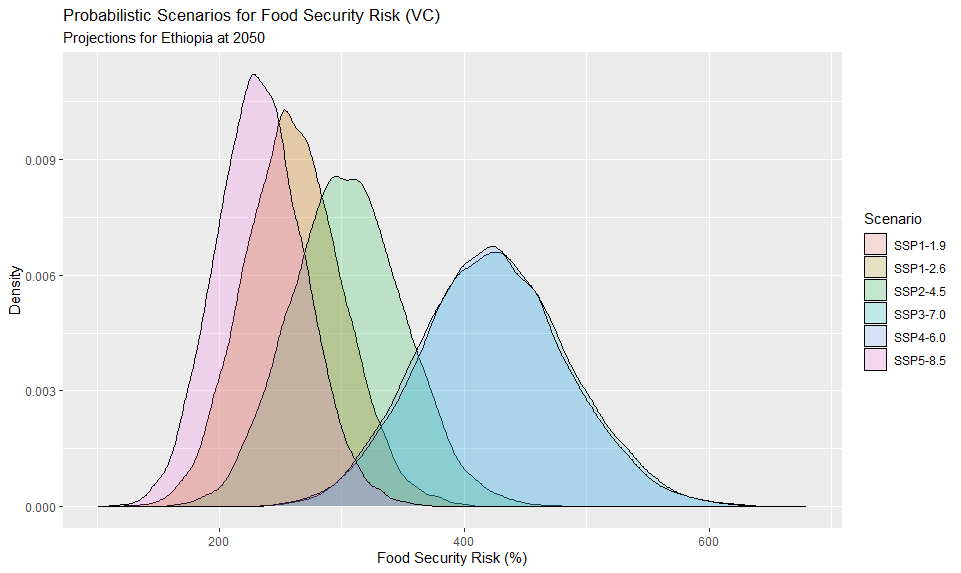}
\caption{Food security risk (VC)}
\end{subfigure}

\caption{Probabilistic projections for Ethiopia at the year 2050 for food security risk and relevant key quantities}\label{prob-proj-ETH}
\end{figure}

\newpage
\section{Extra Material from Section \ref{sec-4}}\label{App-C}

\subsection*{List of data and web sources used for the model fitting task}

\mbox{} 

\begin{table}[ht]\scriptsize
\centering
\begin{tabular}{l|l|l}
	\hline\hline
	\multicolumn{3}{l}{\bf Train Dataset (Time Period: 1990 - 2018)}\\
	\hline
	\bf Driver  & \bf Unit & \bf Data Source\\  
	\hline
	Agricultural land                                & 1000 ha & FAO Database\\
	GDP (per capita)                                & USD (constant 2015)                       & World Bank Database\\
	Labour (total)                                     & 1000 people           & World Bank Database \\ 
	Labour (in agricultural activities)      & \% of total labour   & World Bank Database (ILO estimates)\\ 
		Population                                          & 1000 people            & World Bank Database\\ 
		Food Supply                                       &  Kcal/capita/day     & FAO Database\\ 
		Food Production Quantity                 & tonnes                    & FAO Database\\ 
		Exported food quantity                     & tonnes                     & FAO Database\\ 
		Imported food quantity                     & tonnes                    & FAO Database\\ 
		Precipitation (yearly average)          & mm                    & Climate Knowledge Portal\\ 
		Temperature (yearly average)         & degrees of Celsium   & Climate Knowledge Portal\\ 
		Water Stress Indicator (SDG 6.4.2)                & \% of internal water resources                      & FAO/AQUASTAT Database\\ 
		\hline\hline
	\end{tabular}
	\caption{Online data sources from which data where used for tuning the models for food system capacity stated in Section \ref{tfs-mod}}\label{tab-train-data}
\end{table}

\subsection*{Cobb-Douglas Modelling Approach}

For each country the following set of parametric models (Cobb-Douglas type for both layers) are considered and fitted to data:\\ \\
	(a) Set of Upper Layer Models
	\begin{eqnarray*}
		\left\{
		\begin{array}{ll}
			C_{c,t}^{FS} \sim & \alpha_{C,0} e^{\alpha_{C,1} t}  \left( Q_{c,t}^{Dom} \right)^{\alpha_{C,2}}  \left( Q^{Exp}_{c,t} \right)^{\alpha_{C,3}}  \left( Q^{Imp}_{c,t} \right)^{\alpha_{C,4}} \\
			Q_{c,t}^{Dom} \sim & \alpha_{D,0} e^{\alpha_{D,1} t} \,\,\left( P_{c,t} \right)^{\alpha_{D,2}} \left( I_{c,t-1} \right)^{\alpha_{D,3}} \left( L_{c,t}^{Agr} \right)^{\alpha_{D,4}} \left( W_{c,t} \right)^{\alpha_{D,5}}  \left( A_{c,t}\right)^{\alpha_{D,6}}  \left( T_{c,t} \right)^{\alpha_{D,7}} \\
			W_{c,t} \sim & \alpha_{W,0} e^{\alpha_{W,1} t} \left( P_{c,t} \right)^{\alpha_{W,2}}  \left( I_{c,t-1} \right)^{\alpha_{W,3}}  \left( Q^{Dom}_{c,t} \right)^{\alpha_{W,4}}  \left( A_{c,t} \right)^{\alpha_{W,5}}  \left( T_{c,t} \right)^{\alpha_{W,6}}  \left( Pr_{c,t} \right)^{\alpha_{W,7}} \\
			A_{c,t} \sim & \alpha_{A,0} e^{\alpha_{A,1} t} \,\, \left( P_{c,t} \right)^{\alpha_{A,2}}  \left( I_{c,t-1} \right)^{\alpha_{A,3}}  \left( Q_{c,t-1}^{Dom} \right)^{\alpha_{A,4}} \\
		\end{array} \right.
	\end{eqnarray*}
	(b) Set of Lower Layer Models
	\begin{eqnarray*}
		\left\{
		\begin{array}{ll}		
			Q^{Exp}_{c,t} \sim & \beta_{E,0} e^{\beta_{E,1} t} \,\, \left( P_{c,t} \right)^{\beta_{E,2}}  \left( I_{c,t-1} \right)^{\beta_{E,3}} \\
			Q^{Imp}_{c,t} \sim & \beta_{I,0} e^{\beta_{I,1} t} \,\, \left( P_{c,t} \right)^{\beta_{I,2}}  \left( I_{c,t-1} \right)^{\beta_{I,3}}  \\
			L^{Agr}_{c,t} \sim & \beta_{L,0} e^{\beta_{L,1} t} \,\, \left( P_{c,t} \right)^{\beta_{L,2}}  \left( I_{c,t-1} \right)^{\beta_{L,3}} \left( L_{c,t} \right)^{\beta_{L,4}}  
		\end{array} \right.
	\end{eqnarray*}
following the abbreviations defined in Section \ref{tfs-mod}, for $c=EGY, ETH$ with country-specific parameter vectors ${\bm \alpha}_c := ( {\bm \alpha}_{C,c}, {\bm \alpha}_{D,c}, {\bm \alpha}_{W,c}, {\bm \alpha}_{A,c} )'$ and ${\bm \beta}_c := ( {\bm \beta}_{E,c},  {\bm \beta}_{I,c}, {\bm \beta}_{L,c}  )'$ for the upper and the lower layer models, respectively.

\newpage
\subsection*{The estimated models for Egypt and Ethiopia}

\begin{table}[ht!]\scriptsize
	\centering
	\begin{tabular}{p{3.3cm} |p{1.7cm} p{1cm} p{1.5cm} p{1.4cm} |p{1.25cm} p{1.25cm} p{1.25cm}}
		\hline\hline
		& \multicolumn{4}{c|}{\bf Upper Layer} & \multicolumn{3}{c}{\bf Lower Layer}\\
		\bf Predictor & Food System Capacity & Water Stress & Dom. Food Prod. Qty & Crop Land (used) & Labour (Agr) & Exported Food Qty & Imported Food Qty\\ 
		\hline
		Technology coef.   & 68.4069       &$\,\,$0.2254 &$\,\,$5.6151   & 0.9429 & 2.8004 & 0.0011 & 0.4020\\
		Yearly trend           &  $\,\,0.0160$ &       -0.0010  &$\,\,$0.0045 & 0.0022  & 0.0030 & 0.0296 & 0.0124\\
		Population              &                       &        -0.0799 &$\,\,$0.2275  & 0.1131   & 0.0580 & 1.3062 & 0.5991\\
		Labour (Total)        &                       &                       &                       &               & 0.1580 &               &       \\
		Labour (Agr)           &                       &                      &$\,\,$0.1957   &        &        &        &       \\
		GDP (per capita)    &                       &       -0.0705 &$\,\,$0.1202  & 0.0299 & 0.0676 & 0.8897 & 0.5627\\
		Exported Qty          &-0.0149         &                       &                       &        &        &        &       \\
		Imported Qty          &$\,\,0.0310$  &                       &                       &        &        &        &       \\
		Dom. Food Prod.    &$\,\,0.2419$  &$\,\,$0.2065  &                       & 0.1873 &        &        &       \\
		Water Stress           &                      &                       &$\,\,$0.3373   &        &        &        &       \\
		Crop Land (used)   &                      &$\,\,$0.3869  &$\,\,$0.4858   &        &        &        &       \\
		Temperature (avg) &                     &$\,\,$0.3480  & -0.0483         &        &        &        &       \\
		Precipitation (avg)  &                     &-0.0695         &                        &        &        &        &       \\ \hline
		Explained Deviance ($R^2$) & 99.53\%        & $\,\,$44.56\% & 94.27\%          & 89.21\% & 58.11\% & 92.70\% & 89.48\%\\
		\hline\hline
	\end{tabular}
	\caption{The two-layer model parameter estimates for Egypt using as trainset the available data from period 1990-2018}\label{fsc-mod-EGY}
\end{table}

\begin{table}[ht!]\scriptsize
	\centering
		\begin{tabular}{p{3.3cm} |p{1.7cm} p{1cm} p{1.5cm} p{1.4cm} | p{1.25cm} p{1.25cm} p{1.25cm}}
		\hline\hline
		& \multicolumn{4}{c|}{\bf Upper Layer} & \multicolumn{3}{c}{\bf Lower Layer}\\
		\bf Predictor & Food System Capacity & Water Stress & Dom. Food Prod. Qty & Crop Land (used) & Labour (Agr) & Exported Food Qty & Imported Food Qty\\ 
		\hline
		Technology coef.     & 58.4825 &$\,\,$0.0039&$\,\,$0.0005&$\,\,$0.1796 &$\,\,$1.1033 & 0.0127 & 0.0889\\
		Yearly trend         & 0.0272  &$\,\,$0.0128&$\,\,$0.0177&$\,\,$0.0224 &$\,\,$0.0054 & 0.0235 & 0.0199\\
		Population           &         &$\,\,$0.5546&$\,\,$0.6245&$\,\,$1.1185 &$\,\,$0.2827 & 0.8926 & 0.6915\\
		Labour (Total)       &         &            &            &             &$\,\,$0.5119 &        &       \\
		Labour (Agr)         &         &            &$\,\,$0.4972&             &             &        &       \\
		GDP (per capita)     &         &-0.2472     &$\,\,$0.2818&-0.0599      &     -0.1051 & 0.5334 & 0.5877\\
		Exported Qty         &0.0181   &            &            &             &             &        &       \\
		Imported Qty         &0.0198   &            &            &             &             &        &       \\
		Dom. Food Prod.      &0.2795   &  -0.0900          &            &-0.3790      &             &        &       \\
		Water Stress         &         &            &-0.1111     &             &             &        &       \\
		Crop Land (used)     &         &$\,\,$1.4905&$\,\,$0.2077&             &             &        &       \\
		Temperature (avg)    &         &$\,\,$1.1754&$\,\,$1.7254&             &             &        &       \\
		Precipitation (avg)  &         &-0.8630     &            &             &             &        &       \\
		\hline
		Explained Deviance ($R^2$) & 99.78\% & 95.74\% & 97.80\% & 90.21\% & 99.62\% & 73.62\% & 64.14\%\\
		\hline\hline
	\end{tabular}
	\caption{The two-layer model parameter estimates for Ethiopia using as trainset the available data from period 1990-2018}\label{fsc-mod-ETH}
\end{table}

\subsection*{List of data and models used for the projection task}

\begin{table}[ht!]\scriptsize
	\centering
	\begin{tabular}{l|l|c|l}
		\hline\hline
		\multicolumn{4}{l}{\bf Projections Data Sources \& Models}\\
		\hline
		{\bf Driver} & {\bf Unit} & {\bf Scenario Dependence} & {\bf Data Source/model} \\
		\hline
		GDP (per capita) & USD (constant 2015) & SSP  & MaGE\\	
		Labour (total) & 1000 people & SSP & MaGE\\ 
		Population & 1000 people & SSP & BayesPop + SSP Modeller\\
		Precipitation (yearly average) & mm                            & RCP & Climate Knowledge Portal \\
		Temperature (yearly average) & degrees of Celsium & RCP & Climate Knowledge Portal \\
		\hline\hline
	\end{tabular}
	\caption{Data sources and models used for creating projections on basic socioeconomic and climate drivers for the time period 2019-2050}\label{tab-proj-data}
\end{table}

\newpage
\subsection*{Water-stress projections under SSP-RCP scenarios}

\begin{figure}[ht!]
\centering
\includegraphics[width=5in]{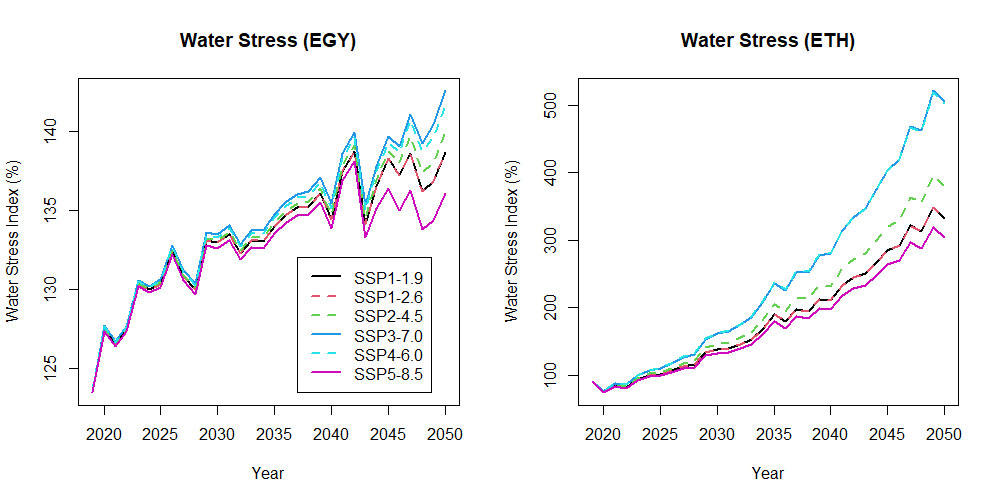}
\caption{Graphical illustration of the water stress index for Egypt and Ethiopia for the time period 2019-2050 under each scenario (median estimates).}\label{fig-wsi-EGY-ETH}
\end{figure}

\bibliographystyle{chicago}
\bibliography{biblio}

\end{document}